\documentclass[twocolumn,aip,nofootinbib,jcp,reprint,amsmath,amssymb]{revtex4-1}

\usepackage{natbib,hyperref}
\usepackage{amsmath}
\usepackage{graphicx}
\usepackage{hyperref} 
\usepackage{dcolumn}
\usepackage{bm}
\usepackage{epsfig,color,xspace,multirow,xr,bbold}
\usepackage[all]{xy}
\usepackage{setspace}
\usepackage{url}
\usepackage{soul}
\usepackage[colorinlistoftodos]{todonotes}
\usepackage{longtable}
\usepackage{tabularx}
\usepackage{adjustbox}
\usepackage{threeparttable} 
\usepackage{natbib,hyperref}
\usepackage{float}
\usepackage{placeins}
\usepackage[utf8]{inputenc}

\usepackage[normalem]{ulem}
\usepackage{xcolor}
\newcommand{\RRef}[1]{Ref.~\onlinecite{#1}}

\newcommand\notsotiny{\@setfontsize\notsotiny\@vipt\@viipt}
\usepackage{soul} 
\begin{document}

\title{
Leveraging the Bias-Variance Tradeoff in Quantum Chemistry for Accurate 
Negative Singlet-Triplet Gap Predictions: A Case for Double-Hybrid DFT
}
\date{\today}

\author{Atreyee Majumdar}
\author{Raghunathan Ramakrishnan}

\email{ramakrishnan@tifrh.res.in}
\affiliation{$^1$Tata Institute of Fundamental Research, Hyderabad 500046, India}

\keywords{
Electronic excited states,
Singlet-Triplet energy gap,
Hund's rule violation,
DFT,
Exchange-Correlation
}

\begin{abstract}
\noindent
Molecules that have been suggested to violate the Hund’s rule, having a first excited singlet state (S$_1$) energetically below the triplet state (T$_1$), are rare. Yet, they hold the promise to be efficient light emitters. Their high-throughput identification demands exceptionally accurate excited-state modeling to minimize qualitatively wrong predictions. We benchmark twelve S$_1$–T$_1$ energy gaps to find that the local-correlated versions of ADC(2) and CC2 excited state methods deliver excellent accuracy and speed for screening medium-sized molecules. Notably, we find that double-hybrid DFT approximations (e.g., B2GP-PLYP and PBE-QIDH) exhibit high mean absolute errors ($>100$ meV) despite very low standard deviations ($\approx10$ meV). Exploring their parameter space reveals that a configuration with 75\% exchange and 55\% correlation, which reduces the mean absolute error to below $5$ meV, but with an increased variance. Using this low-bias parameterization as an internal reference, we correct the systematic error while maintaining low variance, effectively combining the strengths of both low-bias and low-variance DFT parameterizations to enhance overall accuracy. Our findings suggest that low-variance DFT methods, often overlooked due to their high bias, can serve as reliable tools for predictive modeling in first-principles molecular design. The bias-correction data-fitting procedure can be applied to any general problem where two flavors of a method, one with low bias and another with low variance, have been identified {\it a priori}. 
\end{abstract}

\maketitle

\section{Introduction}\label{sec:introduction}

The energy criterion for organic light-emitting diodes (OLEDs) to operate via thermally activated delayed fluorescence (TADF) is that the energy of the first excited singlet state (S$_1$) must be no more than $0.1$ eV higher than that of the triplet state (T$_1$)~\cite{uoyama2012highly,dos2024golden}. 
This condition facilitates thermally-aided reverse intersystem crossing (RISC), enabling the theoretical complete transfer of the T$_1$ population to the emissive state, S$_1$~\cite{chen2018thermally}. 
A derivative of the triangular molecule, heptazine (aka heptaazaphenalene, 7AP), with a central nitrogen (N) atom in an anti-aromatic 12-annulene framework, showed delayed fluorescence without thermal assistance, indicating a negative S$_1$–T$_1$ gap (STG)~\cite{aizawa2022delayed}. 
This mechanism paves the way for designing OLEDs that leverage exothermic delayed fluorescence from 
inverted singlet and triplet excited states (DFIST).
Direct spectroscopic evidence of a negative STG ($-0.047 \pm 0.007$ eV) was obtained for pentaazaphenalene (5AP) using anion photoelectron spectroscopy and fluorescence measurements~\cite{wilson2024spectroscopic}.
Negative STG was also observed in the transient photoluminescence spectrum of dialkylamine-substituted 5AP~\cite{kusakabe2024inverted}.
Computational studies have further suggested that azaphenalenes (APs) with other substitution patterns, as well as the boron (B) analog of cyclazine (1AP),  boraphenalene (1BP), also exhibit negative STGs~\cite{won2023inverted,li2022organic,tuvckova2022origin,dreuw2023inverted,ghosh2022origin,ricci2021singlet,sancho2022violation,loos2023heptazine}.
Non-alternant hydrocarbons and their substituted analogs also show the potential for negative STGs~\cite{garner2023double,terence2023symmetry,garner2024enhanced,blaskovits2024excited,nigam2024artificial}. In particular, substituted analogs of the non-fused bicyclic hydrocarbon have demonstrated negative STGs, which is attributed to through-bond charge-transfer states~\cite{blaskovits2024singlet}. 

A comprehensive search across $12,880$ structurally diverse 
small molecules revealed no exceptions to Hund’s rule, 
indicating that achieving the electronic structure criteria for 
${\rm STG}<0$ requires nontrivial electronic structure and molecular geometry~\cite{majumdar2024resilience}. 
Moreover, the typical magnitudes of negative STGs reported~\cite{loos2023heptazine} based on first-principles calculations 
are smaller than $0.3$ eV, which is similar to the average error associated with popular density functional theory approximations (DFAs).
Consequently, due to the rarity of molecules with negative STGs and the limitations of current quantum chemical approximations, accurately identifying molecules exhibiting ${\rm STG}<0$  via high-throughput screening, with minimal false-positive ({\it i.e.} actual  STG is $>0$ but predicted to be $<0$) and false-negative ({\it i.e.} actual  STG is $<0$ but predicted to be $>0$) predictions, remains challenging. 
The linear-response time-dependent density functional theory (LR-TD-DFT) formalism provides qualitative accuracy---{\it i.e.}, gets the sign of the STG for DFIST candidates correct---only within the double-hybrid (dh) DFT framework containing many-body correlation via the second-order perturbation theory (MP2). 
With the right combination of ingredients, dh-DFT can reliably predict a wide range of molecular properties as benchmarked against higher-level methods or experimental information~\cite{tarnopolsky2008double, goerigk2011efficient,goerigk2014double,goerigk2011efficient}. Nevertheless, selecting suitable exchange-correlation (XC) functionals within the dh-DFT framework remains challenging, especially for novel problems lacking sufficient reference data.

In this study, we revisit the basic mathematics of data fitting to show how to exploit the bias-variance tradeoff in model building to correct
systematic errors in a model which is highly precise ({\it i.e.} low-variance) but has a low accuracy ({\it i.e.} high-bias). 
By exploiting the freedom of parameterization within the dh-DFT formalism, we propose to apply this strategy to model DFIST molecules with 
${\rm STG}<0$. To this end, we evaluate various computational methods for predicting twelve STGs in triangular molecules, benchmarking them against previously reported theoretical best estimates (TBEs) based on composite models~\cite{loos2023heptazine}. 
We examine dh-DFAs exhibiting very small variance for STG predictions and reparametrize them to correct for their bias due to systematic errors. 
For this purpose, we also identify optimal low-bias dh-DFA parameterizations to use as internal references for correcting the predictions of the high-bias, low-variance formalisms. The utility of this internally referenced scaling approach is demonstrated by its ability to predict STGs with low bias and variance, resulting in excellent overall accuracy and precision. We finally discuss the merits and drawbacks of this approach for its further applications.

\section{Dataset and Computational Methods\label{sec:methods}}
\begin{figure}[!htpb]
    \centering
    \includegraphics[width=\linewidth]{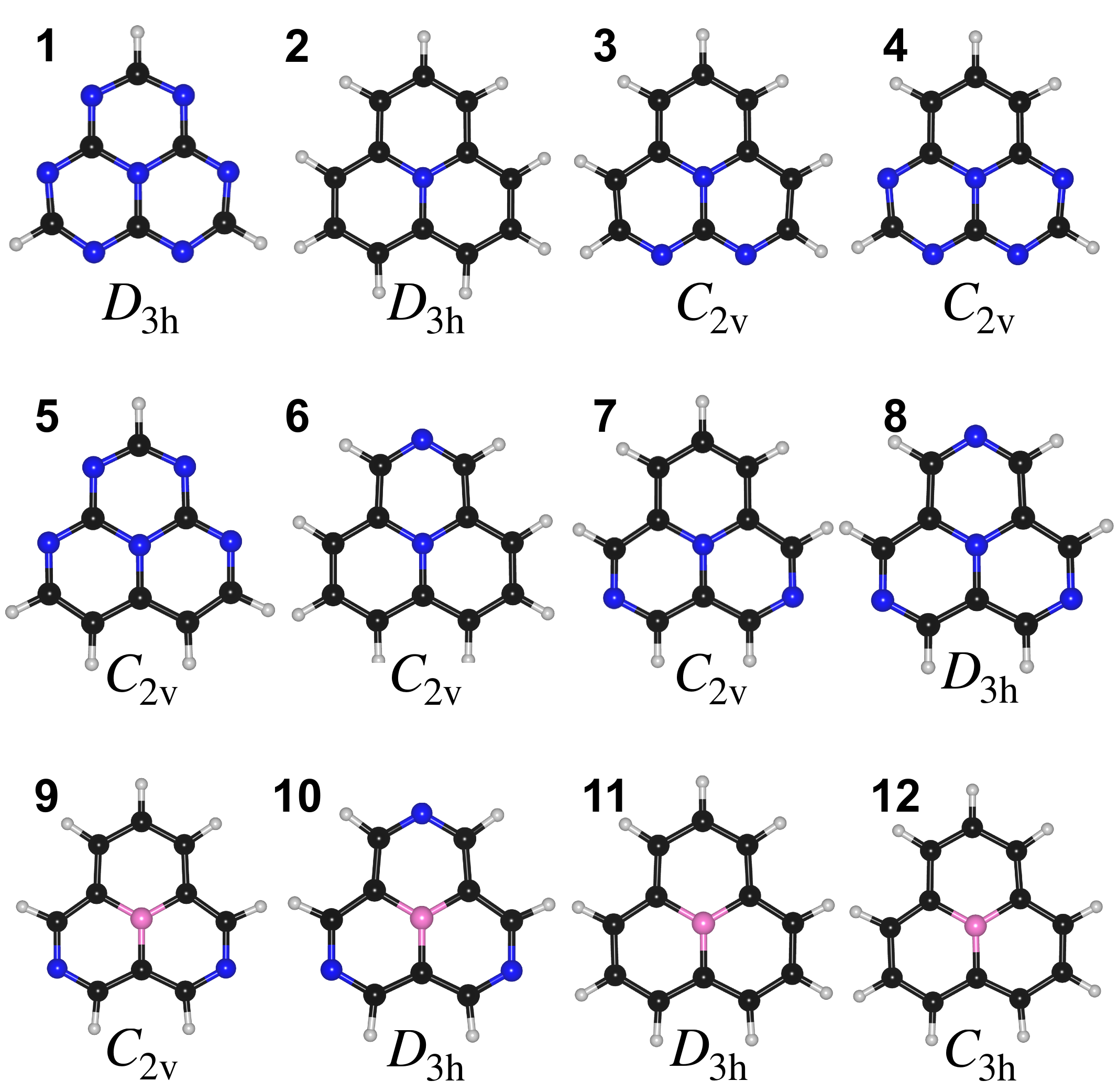}
    \caption{
Nitrogen- and boron-centered [12]-annulene systems reported in
\RRef{loos2023heptazine}.
}
\label{fig:dataset1}
\end{figure}

We consider twelve systems (see Figure~\ref{fig:dataset1}) using equilibrium geometries from \RRef{loos2023heptazine}, determined at the coupled-cluster with singles, doubles, and perturbative triples (CCSD(T)) level, employing the frozen-core approximation and the cc-pVTZ basis set. 
Using these geometries, we calculated the electronic excited state energy levels with various methods in a single-point fashion.
In addition, we collected TBEs of the S$_1$ and T$_1$ energies of the systems shown in Figure~\ref{fig:dataset1} from \RRef{loos2023heptazine} to serve as reference values for benchmarking. 

TBE of STGs reported in \RRef{loos2023heptazine} are based on the composite procedure~\cite{loos2023heptazine,loos2021mountaineering}, where the  
S$_1$ energy is extrapolated as 
CC3/aug-cc-pVTZ + [CCSDT/6-31+G({\it d}) – CC3/6-31+G({\it d})], 
and the T$_1$ energy is extrapolated as
CCSD/aug-cc-pVTZ + [CC3/aug-cc-pVDZ – CCSD/aug-cc-pVDZ]. 
In other words, TBE of S$_1$ energies were computed at the CC3/aug-cc-pVTZ level with a post-CC3 correction via EOM-CCSDT (using the 6-31+G({\it d}) basis set). 
In contrast, the T$_1$ energies were obtained at the EOM-CCSD/aug-cc-pVTZ level with a post-CCSD correction via CC3 using the aug-cc-pVDZ basis set \cite{loos2023heptazine}. 
Although the TBE approach treats S$_1$ and T$_1$ energies using different protocols, both are estimated at the currently available higher levels of theory.


Laplace-transformed, density-fitted, local versions of correlated methods, such as CC2~\cite{freundorfer2010local} and ADC(2)~\cite{schutz2015oscillator}, offer favorable speedups for modeling the electronic excited states of medium-sized organic molecules. 
We denote these methods as L-CC2 and L-ADC(2) and performed calculations using the cc-pVDZ, aug-cc-pVDZ, cc-pVTZ, and aug-cc-pVTZ basis sets, along with the corresponding JKFIT and MP2FIT auxiliary basis sets using Molpro (version 2015.1)~\cite{werner2015molpro}. 
L-CC2 and L-ADC(2) are the same methods denoted as LT-DF-CC2 and LT-DF-ADC(2) elsewhere. 
L-CC2 has demonstrated excellent accuracy in modeling the excited states of BODIPY derivatives~\cite{feldt2021assessment, momeni2016local}.
L-CC2 has been shown to deviate from the canonical variant CC2 by $\le 0.05$ eV, irrespective of the character of the excited states,
including that of charge-transfer states for which CC2's predictions can be erroneous~\cite{kats2006local}. 
The systems studied in this work do not feature charge-transfer characteristics, which is why CC2 predicts accurate STGs in agreement with 
TBEs reported in \RRef{loos2023heptazine}.

For comparison, we collected S$_1$ and T$_1$ energies obtained using the 
linear-response (LR) coupled-cluster methods, EOM-CCSD and CC3, with the aug-cc-pVDZ basis set from \RRef{loos2023heptazine}. 
Additionally, we performed ADC(2) calculations using the cc-pVDZ, cc-pVTZ, aug-cc-pVDZ, and aug-cc-pVTZ basis sets with the resolution-of-the-identity (RI) approximation \cite{vahtras1993integral, kendall1997impact} as implemented in QChem (version 6.0.2) \cite{krylov2013q}. As the excited state energy eigenvalues are typically calculated with iterative eigensolvers, at L-CC(2), L-ADC(2), and ADC(2) levels, we calculated the lowest three singlet and the lowest three triplet roots for sufficient variational convergence of the S$_1$ and T$_1$ states.

We performed LR-TD-DFT calculations using the Tamm--Dancoff approximation (TDA). 
It is known that TDA may not predict oscillator strengths as accurately as the DFT version of random-phase approximation (RPA). However, RPA predictions of triplet energies are sensitive to triplet stability~\cite{rangel2017assessment}. Therefore, we prefer TDA for automated high-throughput calculations of excited-state energies. These energies may be combined with RPA-based oscillator strengths for more detailed analysis.  
Furthermore, previous work has demonstrated that its STG predictions for azaphenalenes are more accurate than those predicted using RPA~\cite{majumdar2024resilience}. The performance of various DFAs for predicting STGs in azaphenalenes has been discussed in prior studies \cite{bhattacharyya2021can, curtis2023building, kunze2024deltadft}. Within the TDA formalism, we calculated the lowest six singlet and the lowest six triplet roots.

This study explores several DFAs 
spanning all levels of the Jacob's ladder \cite{perdew2001jacob}, namely, 
local density approximation (LDA): VWN5 \cite{wilk1981fermi} and PWLDA;
generalized gradient approximation (GGA): BP86, BLYP, PW91 \cite{perdew1992accurate,burke1998derivation}, PBE \cite{perdew1996generalized}, mPWPW, and mPWLYP;
hybrid GGA: B3LYP \cite{stephens1994ab}, PBE0 \cite{adamo1999toward}, X3LYP \cite{su2004accurate}, mPW1PW, mPW1LYP, and BHandHLYP \cite{becke1993new};
meta-GGA: TPSS \cite{perdew2009workhorse}, TPSSh \cite{perdew2009workhorse}, TPSS0 \cite{perdew2009workhorse}, and M06L \cite{zhao2006new}
hybrid meta-GGA: M062X \cite{zhao2008m06};
range-separated hybrids: $\omega$B97, \cite{chai2008systematic} $\omega$B97X, \cite{chai2008systematic}$\omega$B97X-D3, CAM-B3LYP \cite{yanai2004new}, LC-BLYP \cite{tawada2004long}, and LC-PBE \cite{iikura2001long};
double hybrid: $\omega$B97X-2 \cite{chai2009long}, B2PLYP \cite{grimme2006semiempirical}, mPW2PLYP \cite{schwabe2006towards}, B2GP-PLYP \cite{karton2008highly}, PWPB95 \cite{goerigk2011efficient}, PBE-QIDH \cite{bremond2014communication}, PBE0-DH \cite{bremond2011seeking}, $\omega$B2PLYP \cite{casanova2019omegab2plyp}, $\omega$B2GP-PLYP \cite{casanova2019omegab2plyp}, $\omega$B88PP86 \cite{casanova2021time}, $\omega$PBEPP86 \cite{casanova2021time}, SCS/SOS-B2PLYP21 \cite{casanova2021time}, SCS-PBE-QIDH \cite{casanova2021time}, SOS-PBE-QIDH \cite{casanova2021time}, SCS-B2GP-PLYP21 \cite{casanova2021time}, SOS-B2GP-PLYP21 \cite{casanova2021time}, SCS-$\omega$B2GP-PLYP21 \cite{casanova2021time}, SOS-$\omega$B2GP-PLYP21 \cite{casanova2021time}, SCS-RSX-QIDH \cite{casanova2021time}, SOS-RSX-QIDH \cite{casanova2021time}, SCS-$\omega$B88PP86 \cite{casanova2021time}, SOS-$\omega$B88PP86 \cite{casanova2021time}, SCS-$\omega$PBEPP86 \cite{casanova2021time}, SOS-$\omega$PBEPP86 \cite{casanova2021time}, DSD-BLYP \cite{yu2014spin}, DSD-PBEP86 \cite{yu2014spin}, DSD-PBEB95 \cite{yu2014spin}, RSX-QIDH \cite{bremond2018range}, and RSX-0H \cite{bremond2019range}. Here, SCS and SOS denote spin-component-scaled and spin-opposite-scaled, respectively. Such corrections have generally been shown to improve the MP2 correlation component in dh-DFAs.
All DFT calculations were performed using Orca (version 6.0.0) \cite{neese2012orca, neese2018software}.

\section{Bias Correction for Low-Variance Predictors\label{sec:theory}}
Generally, when developing a mathematical model, such as deductive models based on quantum mechanical approximations or inductive models based on data fitting or machine learning, a desired aspect is that the resulting mathematical model is both precise (with low variance) and accurate (with low bias or systematic error). 
However, due to the trade-offs involved in incorporating physical effects while keeping the model computationally tractable, computational chemistry models, particularly DFAs, often exhibit a varying degree of systematic errors across different formulations, such as local density approximation (LDA), generalized gradient approximation (GGA), hybrid-GGA, and so on. This aspect has been highlighted for numerous properties, such as 
binding energies across alkanes~\cite{wodrich2006systematic}, spin-splitting energies~\cite{hughes2011correcting}, chemical reaction energies~\cite{simm2016systematic}, to name a few instances. For atomization or formation energies, DFT-based predictions are often corrected for systematic errors via ``dressed atom corrections'', facilitating decreasing across DFAs while climbing higher on the ``Jacob's ladder of DFAs'' \cite{winget2004enthalpies,grimme2006semiempirical,das2021critical}.


The main goal of this study is to derive formal relations that enable one to 
correct a low-variance, high-bias model ({\it i.e.} precise but inaccurate) 
by correcting it through linear regression using a high-variance, low-bias model as an internal reference. 
We begin by examining the conditions under which such an internally referenced bias correction is effective. 
To the best of our knowledge, the details of this procedure have not been reported elsewhere. However, given the simplicity of the 
technique proposed here, it could be included in a plethora of calibration techniques used in data-fitting~\cite{datafittingwohland}.

\subsection{Definitions} 

We begin by defining the key error metrics and terms used in this study.

\begin{enumerate}
    \item Let $\hat{y}_1$ and $\hat{y}_2$ denote two predictive models (or predictors) for a given property. In this study, they represent two parameterizations of a dh-DFA used for determining STG. Further, $\hat{y}^{\rm true}$ represents the exact theoretical value of the property under consideration.
    The best available finite-data estimate of $\hat{y}^{\rm true}$ is denoted as the reference value, $\hat{y}^{\rm ref}$. The $k$-th prediction 
    from $\hat{y}_1$ and the corresponding reference value are denoted as $\hat{y}_{1,k}$ and $\hat{y}_{{\rm ref},k}$, respectively.
    \item The expected value ({\it i.e.}, mean) of $\hat{y}_1$ is given by $\mathbb{E} [ \hat{y}_1]$.
     The variance of $\hat{y}_1$ around its mean is defined as 
    ${\rm Var}(\hat{y}_1) = \mathbb{E}[\hat{y}_1^2] - \mathbb{E}[\hat{y}_1]^2$.
    The covariance between two predictors measures their linear relationship and is defined as ${\rm Cov}(\hat{y}_1, \hat{y}_2) = 
    \mathbb{E}[\hat{y}_1 \hat{y}_2] - \mathbb{E}[\hat{y}_1] \mathbb{E}[\hat{y}_2]$.
    \item The error metrics are defined as follows. Mean signed error (MSE) is defined as
        ${\rm MSE}(\hat{y}_1) = \mathbb{E} [ \hat{y}_1 - \hat{y}_{\rm ref} ]$, while 
        mean absolute deviation (MAD) is defined as
       ${\rm MAD}(\hat{y}_1) = \mathbb{E} [ | \hat{y}_1 - \hat{y}_{\rm ref} | ]$
        Standard deviation of error (SDE) for a sample of size $N$ is defined as
        ${\rm SDE}(\hat{y}_1) = \sqrt{\frac{1}{N-1} \sum_{k=1}^{N} (\hat{y}_{1,k} - \hat{y}_{{\rm ref},k}-{\rm MSE})^2 }$.
        In the limit of a large sample (\textit{i.e.,} for a population), the square of SDE corresponds to the variance 
        of the error ${\rm Var}(\hat{y}_1-\hat{y}_{\rm ref})$.
        \item The term `bias' refers to a systematic shift in a distribution and requires a reference for comparison. Throughout this study, bias specifically refers to systematic shifts in prediction error $\hat{y}_1 - \hat{y}_{\rm ref}$. 
        Similarly, we distinguish between
        ${\rm Var}(\hat{y}_1)$ and ${\rm Var}(\hat{y}_1 - \hat{y}_{\rm ref})$, where the former indicates spread around the mean value of $\hat{y}_1$ but the latter quantifies the spread of its prediction errors, which is the primary focus of this study.
\end{enumerate}

\subsection{Bias Correction via Ordinary Least Squares Regression\label{biascorrection_OLSR}}  
The bias correction scheme is applicable in cases where two predictors, \( \hat{y}_1 \) and \( \hat{y}_2 \), are available with known bias and variance determined {\it a priori} using reference values, \( \hat{y}^{\rm ref} \).  For example, the bias and variance of the two models can be determined 
using a subset of the entire dataset. 

We assume that \( \hat{y}_1 \) has low variance in prediction errors (\textit{i.e.}, small SDE) but high bias (\textit{i.e.}, large magnitude of MSE), and
\( \hat{y}_2 \) has low bias but high variance in prediction errors.
These conditions are expressed as
\begin{eqnarray}
    | \mathbb{E} \left[ \hat{y}_1 - \hat{y}^{\rm ref}\right] | &>& 0,\, \text{Var}\left( \hat{y}_1  - \hat{y}^{\rm ref}\right) \approx 0\, \text{and}  \nonumber \\ 
     \mathbb{E}\left[ \hat{y}_2 - \hat{y}^{\rm ref} \right] &\approx& 0,\, \text{Var}\left( \hat{y}_2 - \hat{y}^{\rm ref}\right) > 0.
     \label{eq:conditions_y1_y2}
\end{eqnarray}

We define a new estimator \( \hat{y}_1^* \) by linearly scaling \( \hat{y}_1 \)
\begin{eqnarray}
    \hat{y}_1^* & = & a \hat{y}_1 + b,
\end{eqnarray}
where \( a \) and \( b \) are determined via ordinary least squares regression, minimizing the squared error 
\begin{eqnarray}
\min_{a, b} \sum_i \left( a \hat{y}_{1,i} + b - \hat{y}_{2,i} \right)^2
\end{eqnarray}
with respect to \( \hat{y}_2 \), which is used as an internal reference. 
The optimal values of \( a \) and \( b \) are given by
\begin{eqnarray}
a = \frac{\text{Cov}(\hat{y}_1, \hat{y}_2)}{\text{Var}(\hat{y}_1)}; \quad 
b = \mathbb{E}[\hat{y}_2] - a \mathbb{E}[\hat{y}_1].
\label{eq:coeffb}
\end{eqnarray}

\subsection{Error analysis\label{sec:erroranalysis}}
We want to determine whether the bias and variance of \( \hat{y}_1^* \) match those of \( \hat{y}_2 \) and \( \hat{y}_1 \), respectively. 
The mean of the corrected model, \( \hat{y}_1^* \), is
\begin{eqnarray}
\mathbb{E}[\hat{y}_1^*] =
\mathbb{E}[a \hat{y}_1 + b] = a \mathbb{E}[\hat{y}_1] + b.
\end{eqnarray}
Substituting the expression for \( b \) from Eq.~\ref{eq:coeffb}, we find:
\begin{eqnarray}
\mathbb{E}[\hat{y}_1^*] = a \mathbb{E}[\hat{y}_1] + \mathbb{E}[\hat{y}_2] - a \mathbb{E}[\hat{y}_1] = \mathbb{E}[\hat{y}_2].
\end{eqnarray}
Thus, least-squares regression ensures that the mean of \( \hat{y}_1^* \) aligns with that of the reference, the low-bias predictor \( \hat{y}_2 \), thereby correcting the bias. 

The MSE of \( \hat{y}_1^* \) is
\begin{eqnarray}
    \mathbb{E}[\hat{y}_1^* - \hat{y}^{\rm ref}] & = &
    \mathbb{E}[a \hat{y}_1 + b - \hat{y}^{\rm ref}] = 
    \mathbb{E}[a \hat{y}_1 ]  +
    \mathbb{E}[ b ] -
    \mathbb{E}[\hat{y}^{\rm ref}] \nonumber \\
     & = &
     \mathbb{E}[a \hat{y}_1 ]  +
    \left( \mathbb{E}[\hat{y}_2] - a \mathbb{E}[\hat{y}_1] \right) -
    \mathbb{E}[\hat{y}^{\rm ref}]  \nonumber \\ 
    & = & 
    \mathbb{E}[ \hat{y}_2 ] - \mathbb{E}[\hat{y}^{\rm ref}] =
   \mathbb{E}[ \hat{y}_2 - \hat{y}^{\rm ref}].
\end{eqnarray}
Hence, the MSE of \( \hat{y}_1^* \) is identical to that of \( \hat{y}_2 \), confirming that bias is fully transferred from \( \hat{y}_2 \).

The variance of \( \hat{y}_1^* \)  can be derived as
\begin{eqnarray}
\text{Var}(\hat{y}_1^*) = \text{Var}(a \hat{y}_1 + b) = a^2 \text{Var}(\hat{y}_1).
\end{eqnarray}
For \( \hat{y}_1^* \) to preserve the variance of \( \hat{y}_1 \), we require
 \( a^2 = 1 \), which implies \( a = \pm 1 \).  
For \( a \approx 1 \), \( \hat{y}_1^* \) retains the variance of \( \hat{y}_1 \) while aligning its mean with \( \mathbb{E}[\hat{y}_2] \), effectively shifting the prediction distribution. However, in most practical cases, \( |a| \neq 1 \), leading to:
 \( |a| < 1 \) resulting in \( \text{Var}(\hat{y}_1^*) < \text{Var}(\hat{y}_1) \), or \( |a| > 1 \) resulting in \( \text{Var}(\hat{y}_1^*) > \text{Var}(\hat{y}_1) \). 
 The variance of \( \hat{y}_1^* - \hat{y}^{\rm ref} \) is given by:
\begin{eqnarray}
    {\rm Var} \left( \hat{y}_1^* - \hat{y}^{\rm ref} \right) & = &
    {\rm Var} \left( a \hat{y}_1 + b  - \hat{y}^{\rm ref} \right).
\end{eqnarray}
Applying the variance sum formula, ${\rm Var} \left( A+B \right)={\rm Var} \left( A \right)+{\rm Var} \left( B \right)+{\rm Cov} \left( A, B \right)$, we arrive at
\begin{eqnarray}
    {\rm Var} \left( \hat{y}_1^* - \hat{y}^{\rm ref} \right) & = &
    a^2 {\rm Var} \left( \hat{y}_1 - \hat{y}^{\rm ref}/a \right).
\end{eqnarray}
This implies that the variance of \( \hat{y}_1^* - \hat{y}^{\rm ref} \) is not simply \( a^2 \) times the variance of \( \hat{y}_1 - \hat{y}^{\rm ref} \), but is influenced by the distribution of the reference itself.
Thus, only when \( a = 1 \), the variance of \( \hat{y}_1^* - \hat{y}^{\rm ref} \) exactly matches that of \( \hat{y}_1 - \hat{y}^{\rm ref} \). If \( a = -1 \), while the spread 
of \( \hat{y}_1^* \) remains unchanged from that of \( \hat{y}_1 \), the spread of the errors of $\hat{y}_1^*$ will increase.
For general values of \( a \), it is possible that fortuitously,
the variance of prediction errors decreases after scaling.

\begin{table*}[!htpb]
\centering
\caption{Excitation energies of S$_1$ and T$_1$ levels of the twelve benchmark systems FIG~\ref{fig:dataset1},
along with the corresponding STGs, are presented at different theoretical levels using the aug-cc-pVTZ basis set. 
For reference, TBE results are also provided, which are based on composite estimations.  
Error metrics and the reference TBE values are also provided.
MSE: mean signed error,
MAD: mean absolute deviation, and
SDE: standard deviation of the error.
Wherever necessary, the convention for deviation is `Method$-$TBE.' 
All values are in eV.
}
\label{tab:S1T1STG_references}
\footnotesize\addtolength{\tabcolsep}{1.2pt}
\begin{tabular}[t]{l cccc cccc cccc cccc cccc ccc}
\hline
\multicolumn{1}{l}{\#} &
\multicolumn{4}{l}{L-CC2$^a$} &
\multicolumn{4}{l}{CC2$^b$} &
\multicolumn{4}{l}{L-ADC(2)$^a$} &
\multicolumn{4}{l}{ADC(2)$^b$} &
\multicolumn{3}{l}{TBE$^b$} \\
\cline{2-4} \cline{6-8} \cline{10-12} \cline{14-16} \cline{18-20}
\multicolumn{1}{l}{} &
\multicolumn{1}{l}{S$_1$} &
\multicolumn{1}{l}{T$_1$} &
\multicolumn{1}{l}{S$_1$-T$_1$} &
\multicolumn{1}{l}{} &
\multicolumn{1}{l}{S$_1$} &
\multicolumn{1}{l}{T$_1$} &
\multicolumn{1}{l}{S$_1$-T$_1$} &
\multicolumn{1}{l}{} &
\multicolumn{1}{l}{S$_1$} &
\multicolumn{1}{l}{T$_1$} &
\multicolumn{1}{l}{S$_1$-T$_1$} &
\multicolumn{1}{l}{} &
\multicolumn{1}{l}{S$_1$} &
\multicolumn{1}{l}{T$_1$} &
\multicolumn{1}{l}{S$_1$-T$_1$} &
\multicolumn{1}{l}{} &
\multicolumn{1}{l}{S$_1$} &
\multicolumn{1}{l}{T$_1$} &
\multicolumn{1}{l}{S$_1$-T$_1$} \\
\hline
1&$2.759$   &  $2.995$  &  $-0.236$  &&  $2.767$  &  $3.006$  &  $-0.239$  &&  $2.680$  &  $2.934$  &  $-0.254$  &&  $2.675$  &  $2.921$  &  $-0.246$  &&  $2.717$  &  $2.936$  &  $-0.219$  \\
2&$1.043$   &  $1.174$  &  $-0.131$  &&  $1.051$  &  $1.181$  &  $-0.130$  &&  $1.001$  &  $1.138$  &  $-0.137$  &&  $1.001$  &  $1.138$  &  $-0.137$  &&  $0.979$  &  $1.110$  &  $-0.131$  \\
3&$1.606$   &  $1.713$  &  $-0.107$  &&  $1.615$  &  $1.721$  &  $-0.106$  &&  $1.553$  &  $1.666$  &  $-0.113$  &&  $1.551$  &  $1.664$  &  $-0.113$  &&  $1.562$  &  $1.663$  &  $-0.101$  \\
4&$2.220$   &  $2.354$  &  $-0.134$  &&  $2.235$  &  $2.366$  &  $-0.131$  &&  $2.167$  &  $2.300$  &  $-0.133$  &&  $2.159$  &  $2.298$  &  $-0.139$  &&  $2.177$  &  $2.296$  &  $-0.119$  \\
5&$2.162$   &  $2.286$  &  $-0.124$  &&  $2.178$  &  $2.296$  &  $-0.118$  &&  $2.097$  &  $2.228$  &  $-0.131$  &&  $2.098$  &  $2.225$  &  $-0.127$  &&  $2.127$  &  $2.230$  &  $-0.103$  \\
6&$0.896$   &  $0.985$  &  $-0.089$  &&  $0.903$  &  $0.988$  &  $-0.085$  &&  $0.852$  &  $0.951$  &  $-0.099$  &&  $0.851$  &  $0.945$  &  $-0.094$  &&  $0.833$  &  $0.904$  &  $-0.071$  \\
7&$0.755$   &  $0.825$  &  $-0.070$  &&  $0.762$  &  $0.827$  &  $-0.065$  &&  $0.709$  &  $0.787$  &  $-0.078$  &&  $0.708$  &  $0.782$  &  $-0.074$  &&  $0.693$  &  $0.735$  &  $-0.042$  \\
8&$0.612$   &  $0.665$  &  $-0.053$  &&  $0.623$  &  $0.681$  &  $-0.058$  &&  $0.565$  &  $0.630$  &  $-0.065$  &&  $0.565$  &  $0.635$  &  $-0.070$  &&  $0.554$  &  $0.583$  &  $-0.029$  \\
9&$1.321$   &  $1.540$  &  $-0.219$  &&  $1.343$  &  $1.555$  &  $-0.212$  &&  $1.267$  &  $1.488$  &  $-0.221$  &&  $1.274$  &  $1.488$  &  $-0.214$  &&  $1.264$  &  $1.463$  &  $-0.199$  \\
10&$1.578$  &  $1.907$  &  $-0.329$  &&  $1.597$  &  $1.916$  &  $-0.319$  &&  $1.525$  &  $1.846$  &  $-0.321$  &&  $1.529$  &  $1.840$  &  $-0.311$  &&  $1.522$  &  $1.827$  &  $-0.305$  \\
11&$0.828$  &  $1.025$  &  $-0.197$  &&  $0.866$  &  $1.054$  &  $-0.188$  &&  $0.786$  &  $0.997$  &  $-0.211$  &&  $0.799$  &  $0.990$  &  $-0.191$  &&  $0.779$  &  $0.974$  &  $-0.195$  \\
12&$1.246$  &  $1.179$  &  $+0.067$   &&  $1.275$  &  $1.207$  &  $+0.068$   &&  $1.248$  &  $1.177$  &  $+0.071$   &&  $1.262$  &  $1.179$  &  $+0.083$   &&  $1.209$  &  $1.147$  &  $+0.062$   \\
\hline
MSE&$+0.051$  &  $+0.065$  &  $-0.014$  &&  $+0.067$  &  $+0.077$  &  $-0.011$  &&  $+0.003$  &  $+0.023$  &  $-0.020$  &&  $+0.005$  &  $+0.020$  &  $-0.015$  &&    &    &    \\
MAD&$+0.051$  &  $+0.065$  &  $+0.015$  &&  $+0.067$  &  $+0.077$  &  $+0.013$  &&  $+0.017$  &  $+0.023$  &  $+0.022$  &&  $+0.021$  &  $+0.023$  &  $+0.019$  &&    &    &    \\
SDE&$+0.010$  &  $+0.016$  &  $+0.014$   &&  $+0.011$  &  $+0.013$  &  $+0.011$   &&  $+0.021$  &  $+0.018$  &  $+0.013$   &&  $+0.025$  &  $+0.020$  &  $+0.016$   &&   &    &    \\
\hline
\end{tabular}
\begin{tablenotes}
\item \footnotesize{$^a$ This work}
\item \footnotesize{$^b$ From \RRef{loos2023heptazine}}. 
Corrected CC2 and ADC(2)values for System 10 are from \RRef{loos2025correction}.  
 \end{tablenotes}
\end{table*}%

\subsection{Practical Considerations}  
In the above discussion, we designated the predictor with low MSE as an internal reference, \( \hat{y}_2 \), interpreting it as the lower-bias model. 
However, this predictor will also have a low MAD only if all errors, \( \hat{y}_2 - \hat{y}^{\rm ref} \), have the same sign. In particular, if all errors are negative, then \(\text{MAD} = -\text{MSE}\).  
In practice, a vanishing MSE for an approximate model often suggests a multimodal error distribution centered around zero, where positive and negative errors cancel each other out. In such cases, MSE can significantly underestimate the actual bias, making it an unreliable metric. A more robust measure of bias is MAD, which accounts for the magnitude of errors irrespective of sign.  
Thus, in this study, we select \( \hat{y}_2 \) as the predictor with the smallest MAD instead of the smallest MSE. This change does not affect any of the derivations presented above. Notably, the MSE of the scaled low-variance predictor, \( \hat{y}_1^* \), remains close to that of \( \hat{y}_2 \), ensuring bias correction.  Furthermore, since the variance of \( \hat{y}_1^* \) remains approximately the same as that of \( \hat{y}_1 \) (for $a\approx1$) the spread of errors is expected to be narrower. As a result, the MAD of \( \hat{y}_1^* \) will typically be smaller than that of \( \hat{y}_2 \), reflecting a better balance between bias and variance. 

The internally referenced bias correction procedure discussed above is general and can be applied to various scenarios. The main ingredient to using this procedure to solve various quantum chemistry problems lies in obtaining the bias and variance of two formalisms of a method (such as a DFA) through benchmarking against reference data. A more critical aspect is the availability of a suitable \( \hat{y}_1 \) and \( \hat{y}_2 \)  predictors to test the hypothesis presented here. 

It is often the case that for molecular properties, \( \hat{y}_2 \) predictors with low bias (but with high variance) are more commonly reported. For example, 
it is well known that semi-empirical methods, such as PM6 or PM7, are fitted to reference data to reduce bias. Hence, they often exhibit lower systematic errors compared to even the most popular DFAs. For example, for the curated dataset, PPE1694, comprising heats of formation of 1694 molecules, 
PM6 and PM7 achieved mean absolute deviations (MADs) of 4.02 and 3.89 kcal/mol~\cite{das2021critical}. For the same set, B3LYP-hybrid-GGA achieved a higher MAD of 
4.67 kcal/mol. 
However, predictors with low variance and high bias, \( \hat{y}_1 \), are seldom reported, partly because MAD is often preferred as a standard error metric over SDE. Hence, in most studies, the critical information regarding the variance that is better expressed in the form of SDE is missing. 
Further investigation is necessary to fully establish the generality of this approach and determine how it can be transferred and applied to a broader range of molecular systems.

\section{Results and Discussions \label{sec:results}}

\subsection{Performance of L-CC2 and L-ADC(2)}
We begin our discussions with the analysis of the performance of L-CC2 and L-ADC(2), which have not been previously applied to model negative STGs. Table~\ref{tab:S1T1STG_references} presents the excitation energies and STGs of the twelve benchmark systems (shown in Figure~\ref{fig:dataset1}) 
at the CC2 and ADC(2) method from \RRef{loos2023heptazine}, along with results obtained with their local-correlated variants. All values are reported for the aug-cc-pVTZ basis set, which is considered to yield nearly converged S$_1$ and T$_1$ energies for small organic molecules such as the benchmark systems considered here. 
For comparison, we have also collected the TBE values from \RRef{loos2023heptazine}, which we use as reference values to evaluate the accuracies of various
methods. 

As stated in \RRef{loos2023heptazine}, structure 11 in Figure~\ref{fig:dataset1} (with $D_{3\rm h}$ symmetry) corresponds to a 
saddle point on the potential energy surface (PES) at the MP2/6-311G($d$,$p$) level, 
whereas the actual minimum energy geometry exhibits $C_{3\rm h}$ symmetry 
(structure 12 in Figure~\ref{fig:dataset1}). Another study~\cite{majumdar2024influence} found that the high-symmetry forms of structures 6, 7, and 8 are not true minima, while structure 2 is a very shallow potential well at the  CCSD(T)/cc-pVTZ level. 
Overall, four out of twelve structures of DFIST candidate molecules considered in this study are transition states; however, we include them in our benchmarking of STGs due to the availability of TBE-level results and the lack of alternative high-quality data.

\begin{table}[!htpb]
\centering
\caption{L-CC2
energies and L-ADC(2) energies with aug-cc-pVDZ  of the
S$_1$ and T$_1$ states with respect to the S$_0$
ground state along with the singlet-triplet
gap, S$_1$-T$_1$,
of 12 triangular benchmark systems.  
All values are in eV and
\# indicates
compound number as in Figure~\ref{fig:dataset1}.
 }
\label{tab:LCC2_LADC2_AVDZ}
\small\addtolength{\tabcolsep}{1.2pt}
\begin{tabular}[t]{l llll lll}
\hline
\multicolumn{1}{l}{\#}   &
\multicolumn{4}{l}{L-CC2} & 
\multicolumn{3}{l}{L-ADC(2)} \\
\cline{2-4} \cline{6-8} 
\multicolumn{1}{l}{} &
\multicolumn{1}{l}{S$_1$} &
\multicolumn{1}{l}{T$_1$} &
\multicolumn{1}{l}{S$_1$-T$_1$} &
\multicolumn{1}{l}{} &
\multicolumn{1}{l}{S$_1$} &
\multicolumn{1}{l}{T$_1$} &
\multicolumn{1}{l}{S$_1$-T$_1$} \\
\hline 
1&$2.745$ & $2.992$ & $-0.247$ && $2.661$ & $2.916$ & $-0.255$ \\
2&$1.050$ & $1.178$ & $-0.128$ && $1.020$ & $1.154$ & $-0.134$ \\
3&$1.602$ & $1.711$ & $-0.109$ && $1.552$ & $1.672$ & $-0.12$ \\
4&$2.211$ & $2.341$ & $-0.130$ && $2.164$ & $2.296$ & $-0.132$ \\
5&$2.147$ & $2.274$ & $-0.127$ && $2.083$ & $2.221$ & $-0.138$ \\
6&$0.902$ & $0.988$ & $-0.086$ && $0.874$ & $0.976$ & $-0.102$ \\
7&$0.757$ & $0.826$ & $-0.069$ && $0.733$ & $0.813$ & $-0.080$ \\
8&$0.613$ & $0.665$ & $-0.052$ && $0.589$ & $0.652$ & $-0.063$ \\
9&$1.304$ & $1.519$ & $-0.215$ && $1.252$ & $1.472$ & $-0.220$ \\
10&$1.563$ & $1.888$ & $-0.325$ && $1.507$ & $1.825$ & $-0.318$ \\
11&$0.834$ & $1.021$ & $-0.187$ && $0.788$ & $0.987$ & $-0.199$ \\
12&$1.257$ & $1.185$ & $+0.072$ && $1.265$ & $1.180$ & $+0.085$ \\

\hline
MSE   &+0.047 &+0.060 &$-$0.013  && +0.006   & +0.025   & $-$0.019   \\
MAD   &+0.047 &+0.060 &+0.016  && +0.031   & +0.030   & +0.023   \\
SDE   &+0.016 &+0.017 &+0.013  && +0.035   & +0.032   &  +0.017  \\
\hline
\end{tabular}
\end{table}%

Note that for system~10 (in Figure~\ref{fig:dataset1}), STGs at  ADC(2) and CC2 levels were originally reported as $-0.435$ eV and $-0.446$ eV, respectively~\cite{loos2023heptazine}. These values 
were later corrected~\cite{loos2025correction} to $-0.311$ eV and $-0.319$ eV in better agreement with the TBE value of $-0.305$ eV. We considered these
latter values in  Table~\ref{tab:S1T1STG_references}, amounting to 
MSE$|$MAD$|$SDE error metrics as $-0.011|0.013|0.011$ eV and  $-0.015|0.019|0.016$ eV  for
STGs predicted with CC2 and ADC(2) compared to TBE. Even though the errors in STGs are lower for CC2 compared to ADC(2), as seen from Table~\ref{tab:S1T1STG_references}, this is due to a favorable cancellation of errors in CC2. For the individual excitation energies, ADC(2) delivers lower error metrics compared to CC2.

While using the local-correlated variants, the errors for L-CC2 and L-ADC(2) were determined as 
$-0.014|0.015|0.014$ and $-0.020|0.022|0.013$ eV, respectively. These values are similar to those of the canonical methods. For both CC2 and ADC(2), introducing local correlation slightly increases the errors. However, the final errors are still sufficiently low to enable reliable
determination of negative STGs with both L-CC2 and L-ADC(2). 
While we did not perform a speed test to determine the speedups due to local-correlation approximation, we observed L-CC2 and L-ADC(2) when using the smaller aug-cc-pVDZ basis set 
to enable excited-state calculations of molecules larger than the benchmark systems (such as those discussed in Section~\ref{sec:casestudy}), which failed to converge successfully when using RI-based ADC(2) due to CPU memory constraints.


To explore the suitability of L-CC2 and L-ADC(2) for high-throughput data generation for identifying DFIST candidates, we performed calculations with the smaller aug-cc-pVDZ basis set. 
The corresponding results are collected in Table~\ref{tab:LCC2_LADC2_AVDZ}. L-CC2 and L-ADC(2) error metrics MSE$|$MAD$|$SDE determined with the aug-cc-pVDZ basis set 
$-0.013|0.016|0.013$ and $-0.019|0.023|0.017$ eV, respectively, are almost identical to 
the afore-stated values based on the larger aug-cc-pVTZ basis set ($-0.014|0.015|0.014$ and $-0.020|0.022|0.013$ eV, respectively). This ensures that the aug-cc-pVDZ basis set is sufficiently converged for further explorations.

\begin{figure*}[!htpb]
    \centering
    \includegraphics[width=0.9\linewidth]{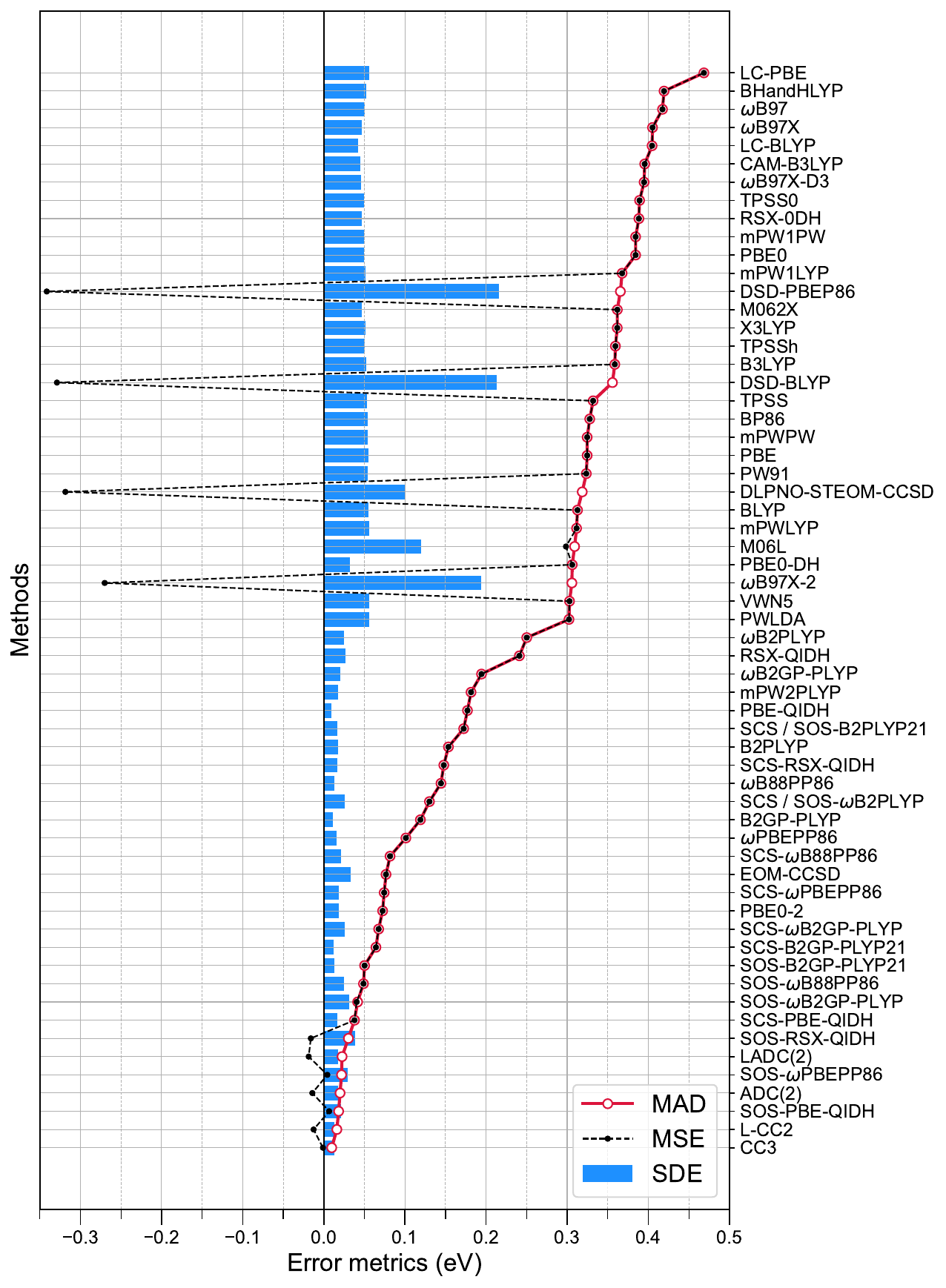}
    \caption{
Error metrics for various methods using the aug-cc-pVDZ basis set to predict 12 theoretical best estimates of STGs for triangular molecules. MAD, MSE, and SDE represent mean absolute deviation, mean signed error, and standard deviation, respectively. Methods are ordered by increasing MAD, with a line included as a visual guide. CC3 and EOM-CCSD results are from~\RRef{loos2023heptazine}. 
}
\label{fig:allmetrics}
\end{figure*}

\subsection{Overall trends across methods}

Figure~\ref{fig:allmetrics} presents an overview of the performance of various theoretical methods, using as a reference the TBEs of STGs collected from \RRef{loos2023heptazine}. 
In this figure, MSE, MAD, and SDE are plotted for all methods, sorted by increasing values of MAD.  In this analysis, we also included CC3/aug-cc-pVDZ results from \RRef{loos2023heptazine}.
For consistency, Figure~\ref{fig:allmetrics} presents results based on the aug-cc-pVDZ basis set, while similar figures for other basis sets are provided in the Supplementary Information (SI).

Among all methods, CC3/aug-cc-pVDZ shows the lowest values for MSE, MAD, and SDE ($-1$, 10, and 13 meV (milli-eV), respectively). As these errors are an order of magnitude
smaller than the typical magnitudes of reported negative STGs, 
CC3/aug-cc-pVDZ can be considered as a computationally affordable reference method. 
The next best performance is observed for L-CC2, with MSE/MAD/SDE values of $-$13/16/13 meV, respectively.   

The MSEs of CC2, L-CC2, ADC(2), L-ADC(2), and CC3 are negative, indicating that these methods will prefer slightly more negative STGs compared to the reference TBE values. Such a systematic trend will result in false-positive predictions of DFIST systems, especially when the magnitudes of the STGs are comparable or smaller than the mean errors of the methods. 
However, if the actual STGs of the DFIST candidate molecules are lower than the MAD of the corresponding method (such as $\lesssim-15$ meV for L-CC2/aug-cc-pVDZ), then predictions can be considered as true-positives. 

EOM-CCSD method is another popular wavefunction method applied to identify
DFIST candidate molecules. The local correlated version of the method DLPNO-STEOM-CCSD
has been very successful in yielding CCSD quality results with attractive speedups. In Table~\ref{table:eomccsd_dlpno}, we present the excited state energies and STGs of the benchmark set obtained with the EOM-CCSD and DLPNO-STEOM-CCSD methods with the aug-cc-pVDZ basis set.

\begin{table}[!hptb]
\centering
\caption{EOM-CCSD
energies and DLPNO-STEOM-CCSD energies with aug-cc-pVDZ  of the
S$_1$ and T$_1$ states with respect to the S$_0$
ground state along with the singlet-triplet
gap, S$_1$-T$_1$,
of 12 triangular benchmark systems.  
All values are in eV and
\# indicates
compound number as in Figure~\ref{fig:dataset1}.
 }
 \label{table:eomccsd_dlpno}
\small\addtolength{\tabcolsep}{1.2pt}
\begin{tabular}[t]{l llll lll}
\hline
\multicolumn{1}{l}{\#}   &
\multicolumn{4}{l}{EOM-CCSD$^a$} & 
\multicolumn{3}{l}{DLPNO-STEOM-CCSD$^b$} \\
\cline{2-4} \cline{6-8} 
\multicolumn{1}{l}{} &
\multicolumn{1}{l}{S$_1$} &
\multicolumn{1}{l}{T$_1$} &
\multicolumn{1}{l}{S$_1$-T$_1$} &
\multicolumn{1}{l}{} &
\multicolumn{1}{l}{S$_1$} &
\multicolumn{1}{l}{T$_1$} &
\multicolumn{1}{l}{S$_1$-T$_1$} \\
\hline 
1&$2.901$ & $3.049$ & $-0.148$ && $2.243$ & $2.930$ & $-0.687$ \\
2&$1.092$ & $1.165$ & $-0.073$ && $0.662$ & $1.093$ & $-0.431$ \\
3&$1.703$ & $1.732$ & $-0.029$ && $1.297$ & $1.649$ & $-0.352$ \\
4&$2.342$ & $2.382$ & $-0.040$ && $1.790$ & $2.275$ & $-0.485$ \\
5&$2.290$ & $2.312$ & $-0.022$ && $1.800$ & $2.181$ & $-0.381$ \\
6&$0.932$ & $0.944$ & $-0.012$ && $0.519$ & $0.888$ & $-0.369$ \\
7&$0.775$ & $0.761$ & $+0.014$ && $0.393$ & $0.685$ & $-0.292$ \\
8&$0.619$ & $0.595$ & $+0.024$ && $0.246$ & $0.527$ & $-0.281$ \\
9&$1.388$ & $1.511$ & $-0.123$ && $0.740$ & $1.362$ & $-0.622$ \\
10&$1.666$ & $1.907$ & $-0.241$ && $0.962$ & $1.747$ & $-0.785$ \\
11&$0.864$ & $0.989$ & $-0.125$ && $0.312$ & $0.850$ & $-0.538$ \\
12&$1.415$ & $1.172$ & $+0.243$ && $0.968$ & $1.020$ & $-0.052$ \\
\hline
MSE   &+0.131 &+0.054 &+0.077  && $-0.374$   &$-0.055$    &  $-0.319$  \\
MAD   &+0.131 &+0.054 &+0.077  && +0.374   & $+0.055$   & +0.319   \\
SDE   &+0.042 &+0.031 & +0.033 && +0.102   & +0.042   & +0.100   \\
\hline
\end{tabular}
\begin{tablenotes}
\item \footnotesize{$^a$ From \RRef{loos2023heptazine}}. 
\item \footnotesize{$^b$ This work}
 \end{tablenotes}
\label{tab:W1}
\end{table}%


Among the wavefunction methods considered in this study, EOM-CCSD is the only method delivering a positive MSE (see Table~\ref{table:eomccsd_dlpno} and also Figure~\ref{fig:allmetrics}). This trend suggests that EOM-CCSD predictions are prone to false-negative predictions in identifying systems with negative STGs, where an actual negative STG can be incorrectly forecasted as positive. 
There is a slight advantage in using EOM-CCSD as its predictions of STGs can serve as upper bounds for negative STGs,
while the actual values can be considered smaller. However, the SDE of EOM-CCSD is $>30$ meV, suggesting that the positive MSE of this method is not systematic. Hence, not all small, positive STGs predicted by EOM-CCSD may be reliably expected to be negative.

DLPNO-STEOM-CCSD has demonstrated to predict the STGs of  multi-resonance TADF compounds in good agreement with EOM-CCSD and ADC(3)~\cite{izu2024reverse},
as well as vertical excitation energies of organic radicals aligning with EOM-CCSD, CC3, and CCSDT~\cite{casanova2024assessment}.
On the otherhand, for the twelve benchmark systems explored here, DLPNO-STEOM-CCSD, exhibits a bias (MAD) exceeding 0.3 eV with an SDE of 0.1 eV (see Table~\ref{table:eomccsd_dlpno} and Figure~\ref{fig:allmetrics}), rendering it unsuitable for applications involving negative STG systems concurring with previous studies~\cite{chanda2024benchmark,chanda2025benchmark}. 
To understand the origin of this error, let us consider System~1 with the TBE values of the excitation energies of 
S$_1$ and T$_1$ states reported as 2.717 eV and 2.936 eV, resulting in an STG of $-0.219$ eV (see Table~\ref{tab:S1T1STG_references}). 
At the EOM-CCSD/aug-cc-pVDZ level,
these values were predicted as 2.901 eV (S$_1$), 3.049 eV (T$_1$), and $-0.148$ eV (STG), see Table~\ref{table:eomccsd_dlpno}. However, with DLPNO-STEOM-CCSD/aug-cc-pVDZ, the same values were predicted as 2.243 eV (S$_1$), 2.930 eV (T$_1$), and $-0.687$ eV (STG), see Table~\ref{table:eomccsd_dlpno}. This comparison suggests that the T$_1$ energies of DLPNO-STEOM-CCSD agree with those of canonical EOM-CCSD, while the former method significantly lowers the S$_1$ energy, resulting in more negative STGs.







Among the DFT methods, only dh-DFAs can predict negative STGs within the LR-TDDFT formalism~\cite{sancho2022violation,casanova2021time}. However, it must be noted that studies have also shown that negative STGs can be predicted using hGGA methods when calculating the S$_1$ and T$_1$ states using the $\Delta$-SCF formalism based on the maximal overlap method~\cite{kunze2024deltadft,chanda2025benchmark}.
In Figure~\ref{fig:allmetrics}, one finds that the dh-DFAs SOS-PBE-QIDH and SOS-$\omega$PBEPP86 have MADs very similar to those of L-CC2, ADC(2), and L-ADC(2). Especially for these two methods, the MSEs are nearly zero, indicating that they intrinsically incorporate bias correction in their parameterization. Further, the dh-DFT approaches B2GP-PLYP and PBE-QIDH  display very small SDEs (Figure~\ref{fig:allmetrics}), indicating that their deviations from the TBE values are primarily systematic. This is a key result that will be explored further in Section~\ref{sec:biasvariance}. The SCS and SOS variants of these methods exhibit smaller MSEs and MADs at the expense of larger SDEs. With few exceptions, the remaining DFAs exhibit significantly larger errors, rendering them unreliable for the first-principles identification of negative STG systems when used within the LR-TDDFT formalism.

\subsection{Bias-variance tradeoff across methods\label{sec:biasvariance}}
\begin{figure}[htp]
    \centering
    \includegraphics[width=\linewidth]{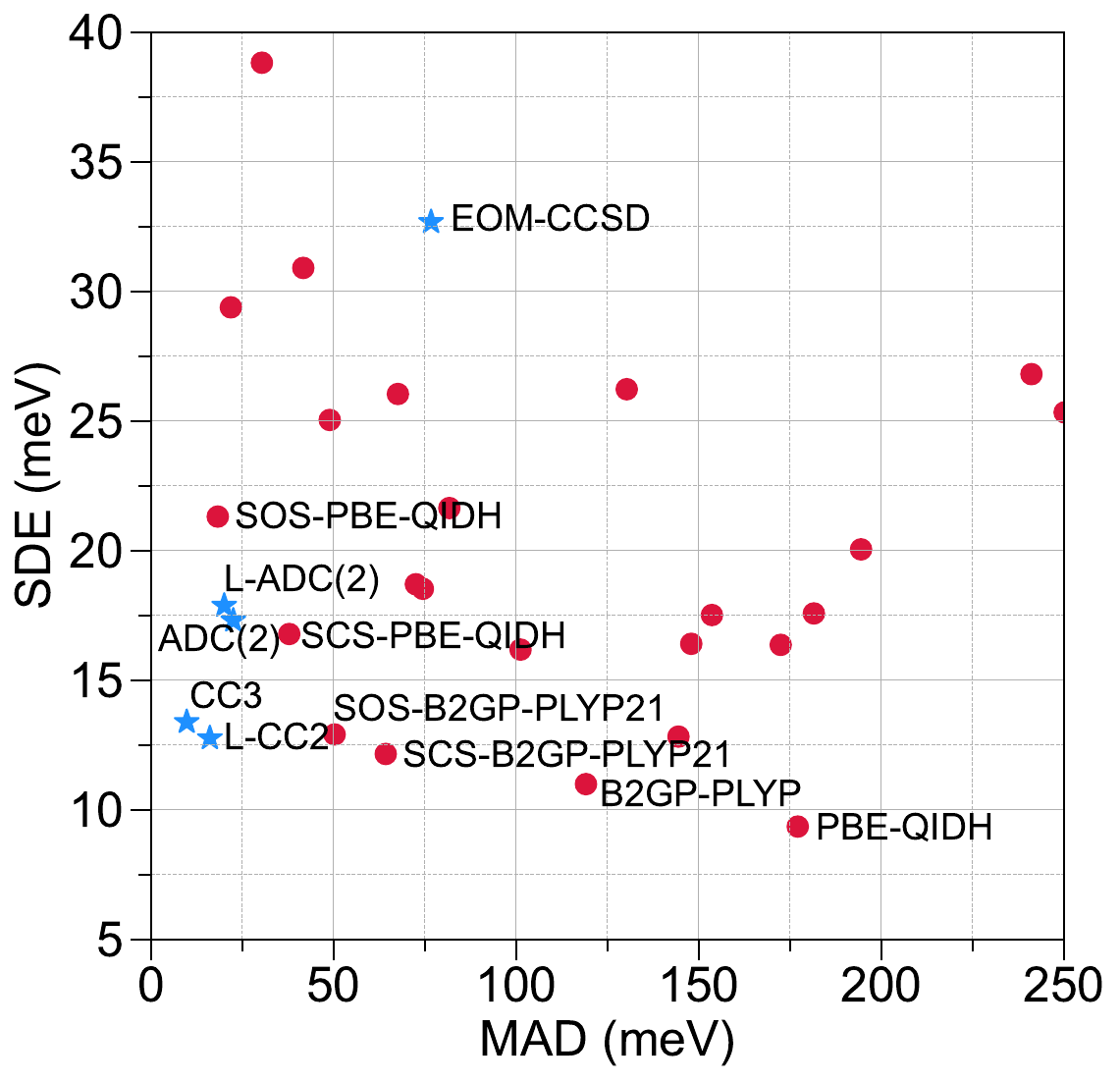}
    \caption{
Bias-variance tradeoff in various methods as a scatterplot of MAD versus SDE for predicting 12 theoretical best estimates of STGs in triangular molecules. Wavefunction methods are marked with blue stars, and DFAs with red circles. For clarity, names are shown only for selected points. All results are based on the aug-cc-pVDZ basis set. CC3 and EOM-CCSD results are from \RRef{loos2023heptazine}.}
\label{fig:scatterbiasvariance}
\end{figure}

While interesting trends in the bias-variance tradeoff of various methods are noticeable in Figure~\ref{fig:allmetrics}, the trend is more clearly revealed in Figure~\ref{fig:scatterbiasvariance}, which displays a scatterplot of MADs and SDEs for various methods (excluding those with high MAD or SDE).
Overall, CC3 and L-CC2 demonstrate superior performance, characterized by low MAD and SDE.
The dh-DFT methods, B2GP-PLYP and PBE-QIDH, exhibit high MADs while maintaining very low SDEs. Such methods are suitable for bias correction, denoted as \( \hat{y}_1 \) in Section~\ref{biascorrection_OLSR}. 
In particular, PBE-QIDH shows an SDE of less than 10 meV, indicating that its predictions are narrowly clustered around the TBE values but shifted by a constant systematic error. Although the MAD of PBE-QIDH exceeds 175 meV, systematic errors are generally easier to correct through linear regression than the non-systematic errors associated with high variance. 

\begin{table*}[!htpb]
\centering
\small 
\caption{For various methods (aug-cc-pVDZ basis set) STGs of benchmark 
systems (Figure~\ref{fig:dataset1}) are presented before and after bias-correction.  
Error metrics and the reference TBE values are also provided.
MSE: mean signed error,
MAD: mean absolute deviation, and
SDE: standard deviation of the error.
Wherever necessary, the convention for deviation is `Method$-$TBE.' 
All values are in eV.
}
\label{tab:scaledresults}
\begin{threeparttable}
\begin{tabular}{l rrr rrr rrr rrr r}
\hline 
\multicolumn{1}{l}{\#} & 
\multicolumn{3}{l}{B2GP-PLYP} & 
\multicolumn{3}{l}{PBE-QIDH} & 
\multicolumn{3}{l}{L-ADC(2)} & 
\multicolumn{3}{l}{L-CC2} & 
\multicolumn{1}{l}{TBE$^e$} \\
\cline{2-3}\cline{5-6}\cline{8-9}\cline{11-12} 
\multicolumn{1}{l}{} & 
\multicolumn{1}{l}{} & \multicolumn{1}{l}{corr.$^a$} & &
\multicolumn{1}{l}{} & \multicolumn{1}{l}{corr.$^b$} & &
\multicolumn{1}{l}{} & \multicolumn{1}{l}{corr.$^c$} & &
\multicolumn{1}{l}{} & \multicolumn{1}{l}{corr.$^d$} & \\
\hline 
 1 & $-$0.095& $-$0.220&& $-$0.035& $-$0.215&& $-$0.255& $-$0.229&& $-$0.247& $-$0.229&& $-$0.219\\
 2 & $-$0.007& $-$0.126&& 0.037& $-$0.141&& $-$0.134& $-$0.116&& $-$0.128& $-$0.116&& $-$0.131\\
 3 & 0.020& $-$0.098&& 0.081& $-$0.095&& $-$0.120& $-$0.103&& $-$0.109& $-$0.098&& $-$0.101\\
 4 & $-$0.014& $-$0.134&& 0.062& $-$0.115&& $-$0.132& $-$0.114&& $-$0.130& $-$0.118&& $-$0.119\\
 5 & 0.005& $-$0.114&& 0.082& $-$0.094&& $-$0.138& $-$0.119&& $-$0.127& $-$0.115&& $-$0.103\\
 6 & 0.042& $-$0.074&& 0.094& $-$0.082&& $-$0.102& $-$0.086&& $-$0.086& $-$0.076&& $-$0.071\\
 7 & 0.070& $-$0.044&& 0.122& $-$0.053&& $-$0.080& $-$0.065&& $-$0.069& $-$0.059&& $-$0.042\\
 8 & 0.081& $-$0.033&& 0.132& $-$0.043&& $-$0.063& $-$0.049&& $-$0.052& $-$0.043&& $-$0.029\\
 9 & $-$0.080& $-$0.204&& $-$0.017& $-$0.196&& $-$0.220& $-$0.197&& $-$0.215& $-$0.199&& $-$0.199\\
10 & $-$0.173& $-$0.303&& $-$0.126& $-$0.308&& $-$0.318& $-$0.289&& $-$0.325& $-$0.304&& $-$0.305\\
11 & $-$0.049& $-$0.171&& $-$0.005& $-$0.184&& $-$0.199& $-$0.177&& $-$0.187& $-$0.172&& $-$0.195\\
12 & 0.177& 0.070&& 0.246& 0.074&& 0.085& 0.090&& 0.072& 0.075&& 0.062\\
\hline
MSE & 0.119& 0.000&& 0.177& 0.000&& $-$0.019& 0.000&& $-$0.013& 0.000&&\\
MAD & 0.119& 0.007&& 0.177& 0.008&& 0.023& 0.014&& 0.016& 0.010&&\\
SDE & 0.011& 0.009&& 0.009& 0.009&& 0.017& 0.016&& 0.013& 0.012&&\\
\hline 
\end{tabular}
\begin{tablenotes}
\item \footnotesize{$^a$ Corrected with ${\rm slope}=1.0654$ and ${\rm intercept}=-0.1190$}
\item \footnotesize{$^b$ Corrected with ${\rm slope}=1.0284$ and ${\rm intercept}=-0.1787$}
\item \footnotesize{$^c$ Corrected with ${\rm slope}=0.9409$ and ${\rm intercept}=-0.0103$}
\item \footnotesize{$^d$ Corrected with ${\rm slope}=0.9546$ and ${\rm intercept}=-0.0065$}
\item \footnotesize{$^e$ From \RRef{loos2023heptazine}}
 \end{tablenotes}
\end{threeparttable}
\end{table*}



Among the local correlated methods, L-CC2 exhibits MAD and SDE values close to those of CC3, while L-ADC(2) shows a slightly larger SDE than L-CC2 (Figure~\ref{fig:scatterbiasvariance}). 
In contrast, the dh-DFT methods B2GP-PLYP and PBE-QIDH display SDEs that are smaller than L-CC2’s, but their MADs are larger by about an order of magnitude. 
To explore the potential of such low-variance, high-bias methods, we applied a least-squares correction to the STG predictions of B2GP-PLYP, PBE-QIDH, L-ADC(2), and L-CC2 using TBE values as the reference. These results are collected in Table~\ref{tab:scaledresults}. For all methods, the MSEs become zero after linear correction as demonstrated in Section~\ref{sec:erroranalysis}. 

L-ADC(2) and L-CC2 exhibit MAD$|$SDEs of $0.023|0.017$ and $0.016|0.013$ eV, respectively, which slightly improve to 
$0.014|0.016$ and $0.010|0.012$ eV upon linear corrections referenced to the TBE values. 
The dh-DFT methods B2GP-PLYP and PBE-QIDH, which initially exhibit MAD/SDEs of $0.119|0.011$ and $0.177|0.009$ eV, retain their SDE upon linear correction, but their MADs drop to less than 0.01 eV. 
This implies that the errors in the STGs predicted by these dh-DFT methods are predominantly systematic in nature.

\begin{table*}[!htpb]
\centering
\caption{Error metrics for predicting twelve values of
S$_1$ and T$_1$ energetics of triangular molecules reported in \RRef{loos2023heptazine}. 
Values are reported compared to the theoretical best estimates (TBE) from \RRef{loos2023heptazine}.
In all cases, the basis set is aug-cc-pVDZ, and the geometries are from \RRef{loos2023heptazine}. 
MSE: mean signed error,
MAD: mean absolute deviation, and
SDE: standard deviation of the error.
Wherever necessary, the convention for deviation is `Method$-$TBE.' 
 All values are in eV.
 }
 \label{tab:tablescssos}
 \begin{threeparttable}
\small\addtolength{\tabcolsep}{1.2pt}
\begin{tabular}[t]{l l rr rr r rr}
\hline
\multicolumn{1}{l}{Method}   &
\multicolumn{1}{l}{Energy} & 
\multicolumn{1}{l}{Slope} & 
\multicolumn{1}{l}{Intercept} & 
\multicolumn{2}{l}{Before correction} & 
\multicolumn{1}{l}{} & 
\multicolumn{2}{l}{After correction} \\ 
\cline{5-6}\cline{8-9} 
\multicolumn{4}{l}{}&
\multicolumn{1}{l}{MAD} & 
\multicolumn{1}{l}{SDE} &
\multicolumn{1}{l}{} & 
\multicolumn{1}{l}{MAD} &
\multicolumn{1}{l}{SDE} \\
\hline 
B2GP-PLYP           &S$_1$ &1.0350 &$-$0.2382 &0.184& 0.030 &&0.012& 0.020\\
                    &T$_1$ &1.0294 &$-$0.1105 &0.065& 0.031 &&0.018& 0.024\\
                     & STG&1.0654 &$-$0.1190 &0.119 &0.011&& 0.007& 0.009\\
SOS-B2GP-PLYP21     &S$_1$ &1.0164 &$-$0.2574 &0.231& 0.015 &&0.008& 0.011\\
                    &T$_1$ &1.0048 &$-$0.1889 &0.181& 0.019 &&0.013& 0.018\\
                     & STG&0.9167 &$-$0.0561 &0.050 &0.013&& 0.007& 0.010\\
SCS-B2GP-PLYP21     &S$_1$ &1.0086 &$-$0.2948 &0.281& 0.013 &&0.009& 0.012\\
                    &T$_1$ &1.0000 &$-$0.2164 &0.216& 0.020 &&0.015& 0.020\\
                     & STG&0.9246 &$-$0.0685 &0.064 &0.012&& 0.008& 0.009\\
PBE-QIDH            &S$_1$ &0.9627 &$-$0.2158 &0.277& 0.034 &&0.013& 0.023\\
                    &T$_1$ &0.9693 &$-$0.0514 &0.100& 0.033 &&0.017& 0.025\\
                     & STG&1.0284 &$-$0.1787 &0.177 &0.009&& 0.008& 0.009\\
SOS-PBE-QIDH        &S$_1$ &1.0120 &$-$0.1281 &0.110& 0.012 &&0.008& 0.009\\
                    &T$_1$ &0.9889 &$-$0.0862 &0.104& 0.017 &&0.012& 0.016\\
                     & STG&0.8423 &$-$0.0246 &0.018 &0.021&& 0.010& 0.012\\
SCS-PBE-QIDH        &S$_1$ &1.0091 &$-$0.1479 &0.134& 0.011 &&0.007& 0.009\\
                    &T$_1$ &0.9907 &$-$0.0818 &0.096& 0.018 &&0.013& 0.017\\
                     & STG&0.8815 &$-$0.0477 &0.038 &0.017&& 0.009& 0.011\\
\hline
\end{tabular}
\end{threeparttable}
\end{table*}%

\subsection{Internally referenced bias-variance correction for double-hybrid DFT}
While one can use L-CC2, CC3, or TBE values to correct PBE-QIDH predictions through linear scaling, we further investigated whether the low-variance DFT methods, B2GP-PLYP and PBE-QIDH, can be corrected to minimize their systematic errors (or intrinsic bias) discussed above. The PBE-QIDH DFA is defined as follows~\cite{bremond2014communication} 
\begin{eqnarray} E_{\rm xc}^{\rm PBE-QIDH} \left[ \rho \right] & = & a_x E_{x}^{\rm HF} + \left( 1-a_x\right) E_{x}^{\rm PBE} + a_c E_{c}^{\rm MP2}\nonumber \\
&& + \left( 1-a_c\right) E_{c}^{\rm PBE}. 
\label{eq:PBEQIDH} 
\end{eqnarray} 
Here, the coefficients $a_x$ and $a_c$ control the fraction of 
exact exchange (via Hartree--Fock) and MP2-level correlation, respectively. 
In the standard formulation, PBE-QIDH ($a_x=0.69$ and $a_c=0.33$)\cite{sandoval2023excitation} belongs to a family of dh-DFAs with $a_c=a_x^3$ delivering low errors for STG \cite{derradji2024searching}. 

Similarly, B2GP-PLYP is based on the GGA-B88 exchange functional combined with LYP-GGA and VWN3-LDA correlation functionals 
\begin{eqnarray} E_{\rm xc}^{\rm B2GP-PLYP} \left[ \rho \right] & = & a_x E_{x}^{\rm HF} + \left( 1-a_x\right) E_{x}^{\rm B88}  + a_c E_{c}^{\rm MP2}\nonumber \\ 
&&  + \left( 1-a_c\right) \left( E_{c}^{\rm VWN3} + E_{c}^{\rm LYP} \right) 
\label{eq:B2GPPYLP} 
\end{eqnarray} 
with standard mixing coefficients $a_x=0.65$ and $a_c=0.36$ \cite{kozuch2010dsd,goerigk2009computation,pantazis2019assessment}.

\begin{figure*}[!htpb]
    \centering
\includegraphics[width=1.0\linewidth]{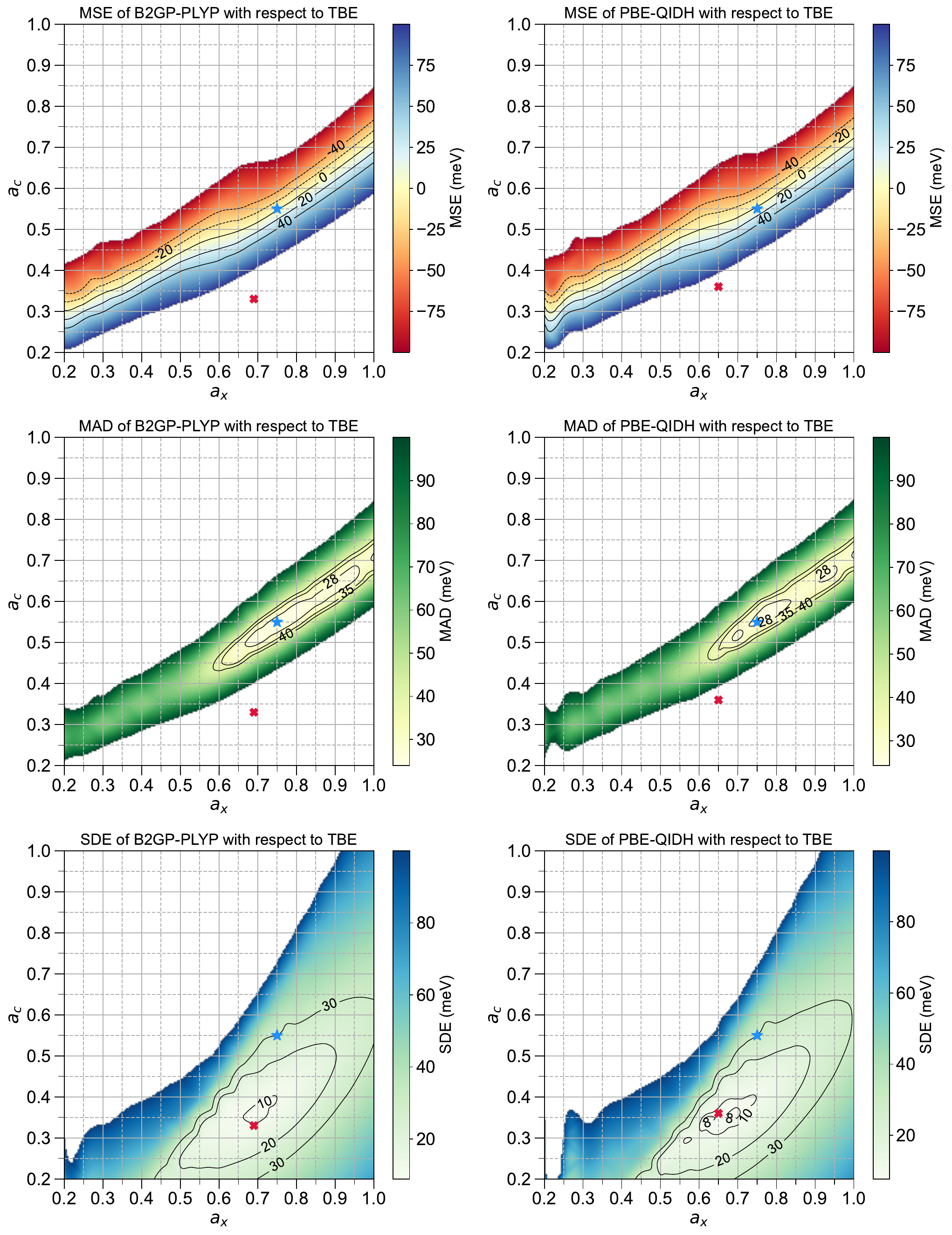}
    \caption{
Dependence of prediction errors of dh-DFT with
exchange and correlation mixing parameters $a_x$ and $a_c$ 
(See Eq.~\ref{eq:PBEQIDH} and Eq.~\ref{eq:B2GPPYLP}).
MSE, MAD and SDE (in meV) are shown for B2GP-PLYP (left panels) and 
PBE-QIDH (right panels) compared to TBE. 
Values of $a_x$ and $a_c$ were sampled in steps of 0.05 and
the surfaces were smoothened using cubic functions. 
In all plots, the default mixing coefficients are marked 
by a red cross---(0.65, 0.36) for  B2GP-PLYP and (0.69, 0.33) 
for PBE-QIDH---and
coefficients that minimize MAD are marked 
by a blue star---(0.75, 0.55) for both B2GP-PLYP and PBE-QIDH.
}
\label{fig:scanalphabeta}
\end{figure*}

Prior studies have shown that the $a_x$-$a_c$ parameter space of B2GP-PLYP exhibits a continuous range of optimal parameters depending on the target property \cite{karton2008highly,goerigk2011efficient}. For instance, Tarnopolsky \textit{et al.} demonstrated variations in the SDE for thermochemistry and reaction barriers as a function of $a_x$ and $a_c$, and proposed the B2T-PLYP and B2K-PLYP df-DFAs with ($a_x=0.60$, $a_c=0.31$) and ($a_x=0.72$, $a_c=0.40$), respectively \cite{tarnopolsky2008double}. 
However, while these studies focused on a single error metric across 
different properties, mainly emphasizing variance because bias in thermochemistry is 
corrected through quasi-atomic corrections\cite{winget2004enthalpies,grimme2006semiempirical,das2021critical}, the bias-variance trade-off for a single property, such as the STG, has not been thoroughly examined.

Using B2GP-PLYP and PBE-QIDH DFAs, we scanned the $a_x$-$a_c$ space and computed error metrics for all 12 benchmark STGs using TBE as the reference. 
As shown in Figure~\ref{fig:scanalphabeta}, both dh-DFAs exhibit low SDEs for the default mixing parameters, and a continuous range of ($a_x$, $a_c$) combinations yielding zero MSE (Figure~\ref{fig:scanalphabeta}, top-most panels). 
Some calculations failed to converge when $a_x$ and $a_c \approx 0$; these are not shown in Figure~\ref{fig:scanalphabeta}. 
However, the minimum MAD is achieved at $a_x=0.75$ and $a_c=0.55$, with both DFAs delivering MADs of approximately 25 meV compared to TBE. 
We denote these optimized parameterizations as B2GP-PLYP (75,55) and PBE-QIDH (75,55). 
To assess the variation of optimal $a_x$ and $a_c$ across error metrics for S$_1$ and T$_1$ energies separately, we performed a comparative analysis (Table~S1 in the SI) using PBE-QIDH. Notably, increased accuracy in S$_1$ energies is observed when both mixing parameters approach 1 (with $a_x=0.90$ and $a_c=0.85$ yielding a MAD of 68 meV), while the lowest MAD for T$_1$ (22 meV) is obtained for $a_x=0.5$ and $a_c=0.15$.

\begin{table}[!hbpt]
\centering
\caption{STGs of benchmark systems shown in Figure~\ref{fig:dataset1}
calculated with double-hybrid DFT methods using the aug-cc-pVDZ basis set.
The error metrics MSE (mean signed error), MAD (mean absolute deviation), and SDE (standard deviation of the error) are with respect to theoretical best estimates given in Table~\ref{tab:scaledresults}. All values are in eV.}
\label{tab:internalreferencescaling}
\begin{threeparttable}
\begin{tabular}{l rrr r rrr}
\hline 
\multicolumn{1}{l}{\#} & 
\multicolumn{3}{l}{B2GP-PLYP} & 
\multicolumn{3}{l}{PBE-QIDH}\\
\cline{2-4}  \cline{6-8} 
  & \multicolumn{1}{l}{(65, 36)} & \multicolumn{1}{l}{(75, 55)} & \multicolumn{1}{l}{scaled$^a$} &  &  
    \multicolumn{1}{l}{(69, 33)} & \multicolumn{1}{l}{(75, 55)} & \multicolumn{1}{l}{scaled$^b$}\\
\hline 
1 & $-$0.095 & $-$0.245 & $-$0.231 & & $-$0.035 & $-$0.239 & $-$0.216 \\
2 & $-$0.007 & $-$0.102 & $-$0.130 & & 0.037 & $-$0.094 & $-$0.138 \\
3 & 0.020 & $-$0.104 & $-$0.099 & & 0.081 & $-$0.096 & $-$0.089 \\
4 & $-$0.014 & $-$0.178 & $-$0.138 & & 0.062 & $-$0.170 & $-$0.110 \\
5 & 0.005 & $-$0.153 & $-$0.116 & & 0.082 & $-$0.144 & $-$0.088 \\
6 & 0.042 & $-$0.064 & $-$0.074 & & 0.094 & $-$0.056 & $-$0.075 \\
7 & 0.070 & $-$0.034 & $-$0.041 & & 0.122 & $-$0.024 & $-$0.045 \\
8 & 0.081 & $-$0.016 & $-$0.029 & & 0.132 & $-$0.007 & $-$0.034 \\
9 & $-$0.080 & $-$0.227 & $-$0.214 & &$-$0.017 & $-$0.222 & $-$0.197 \\
10 & $-$0.173 & $-$0.299 & $-$0.321 &  &$-$0.126 & $-$0.295 & $-$0.316 \\
11 & $-$0.049 & $-$0.154 & $-$0.178 &  &$-$0.005 & $-$0.148 & $-$0.183 \\
12 & 0.177 & 0.086 & 0.082 &  & 0.246 & 0.095 & 0.091 \\
\hline
MSE  & 0.119   & $-$0.003   & $-$0.003  &  &  0.177     &  0.004     &  0.004      \\
MAD  & 0.119   & 0.025   & 0.010  &  &  0.177     &  0.027     &  0.009      \\
SDE  & 0.011  & 0.030   & 0.012  &  &  0.009     &  0.030     &  0.011      \\
\hline 
\end{tabular}
\begin{tablenotes}
 \item \footnotesize{$^a$ Calculated by scaling B2GP-PLYP\,(65,36) values using 
 slope=$1.1521$ and intercept=$-0.1220$} 
 \item \footnotesize{$^b$ Calculated by scaling PBE-QIDH\,(69,33) values using 
 slope=$1.0940$ and intercept=$-0.1780$}
 \end{tablenotes}
\end{threeparttable}
\end{table}

\begin{figure*}[!htpb]
    \centering
    \includegraphics[width=\linewidth]{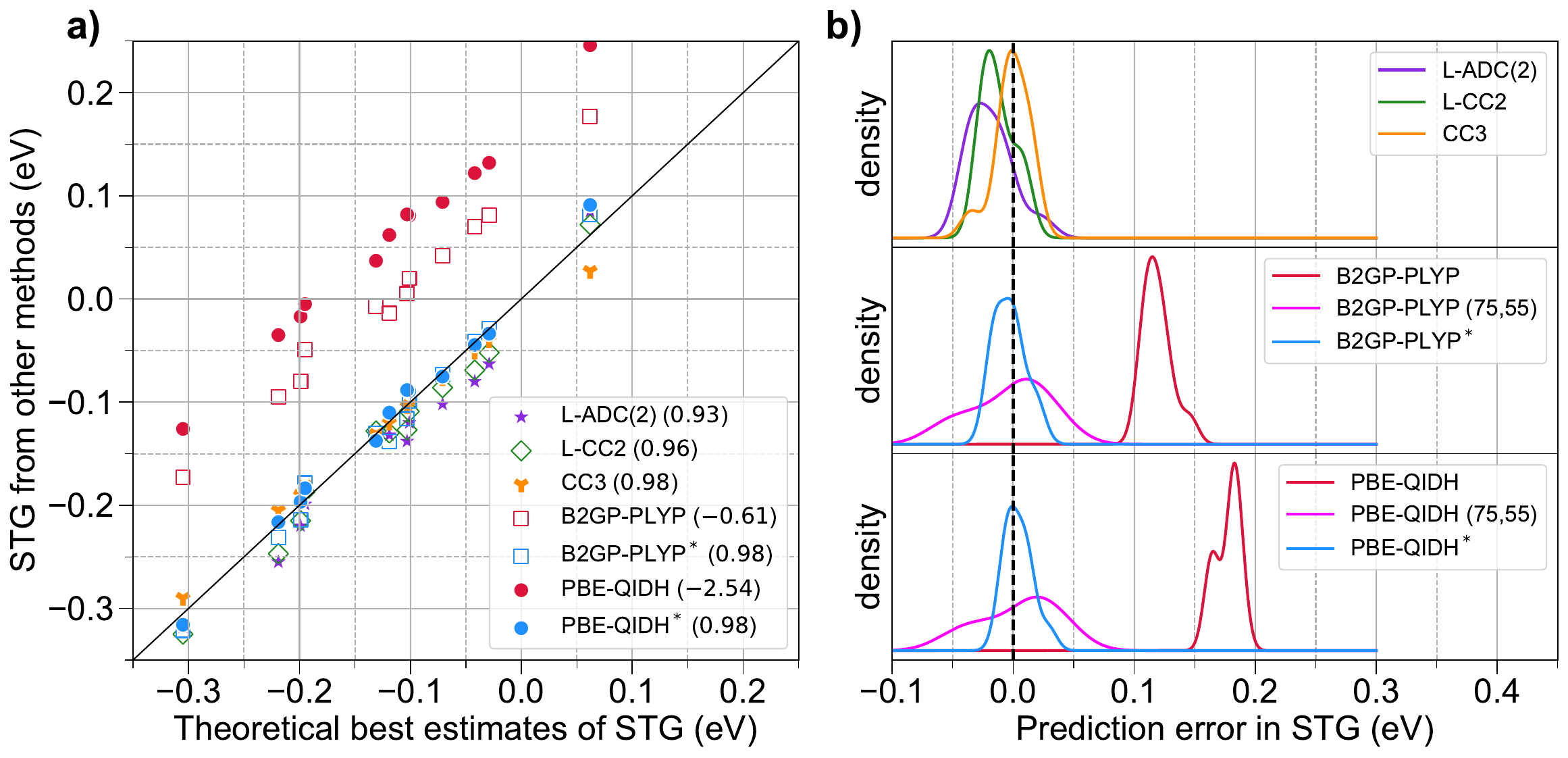}
    \caption{
Comparison of 12 STGs of benchmark systems across methods:
a) Scatterplot of values predicted by various methods (with the aug-cc-pVDZ basis set)
versus theoretical best estimates (TBE). TBE and CC3 results are from \RRef{loos2023heptazine}.
b) Distribution of errors in predicted STGs relative to TBE 
shown as smooth densities. 
}
\label{ref:scatter_errordensity}
\end{figure*}

To apply the scaling protocol discussed in Section~\ref{sec:theory}, the default B2GP-PLYP and PBE-QIDH parameterizations serve as the low-variance estimators ($y_1$ in Eq.~\ref{eq:conditions_y1_y2}), while the (75,55) variants, which exhibit the lowest MAD, serve as the reference low-bias estimators ($y_2$ in Eq.~\ref{eq:conditions_y1_y2}). 
We then determine the slope and intercept via linear regression to 
correct the bias, and denote the corrected predictions as B2GP-PLYP$^*$ and PBE-QIDH$^*$. 

Table~\ref{tab:internalreferencescaling} presents the STG values for the 12 benchmark systems predicted by the bias-corrected dh-DFAs. 
Importantly, the scaling procedure does not require any additional high-level reference data to identify the low-bias or low-variance domain in the parameter space of dh-DFAs. 
The bias-corrected predictions, which simultaneously reflect the small SDE of the base dh-DFAs and the low MAD of the internal references, achieve MAD/SDE values of 10/12 and 9/11~meV for 
B2GP-PLYP$^*$ and PBE-QIDH$^*$, respectively, with respect to TBE.

The overall correlation between the predicted values and TBE is illustrated in Figure~\ref{ref:scatter_errordensity}. Due to the large bias in the original dh-DFT predictions as revealed in Figure~\ref{ref:scatter_errordensity}a, their Pearson correlation coefficients are strongly negative ($\rho=-0.61$ for B2GP-PLYP and $\rho=-2.45$ for PBE-QIDH). 
Upon applying the internally referenced scaling, the correlation improves markedly to $\rho=0.98$ for both DFAs, with all 12 values aligning more closely with the TBE values than those from L-ADC(2), L-CC2, or even CC3. 
The overall performance of the scaling procedure stems from the exceptionally small variance of the dh-DFAs with the default parametrization and the low MAD achieved by the (75,55) parameterizations.

\subsection{Application to heptazine derivatives\label{sec:casestudy}}
To probe the applicability of our bias-correction strategy, we focused on four DFIST candidates proposed in \RRef{aizawa2022delayed}. At the T$_1$ geometry, the STGs of HzTFEX$_2$, HzPipX$_2$, HzTFEP$_2$, and HzTFET$_2$ at the EOM-CCSD, ADC(2), L-CC2, and SCS-ADC(2) levels exhibit smaller magnitudes\cite{aizawa2022delayed} than those of the twelve triangular systems discussed above. This difference is attributed to the change in geometry, as STGs generally increase when moving away from the ground state minimum (S$_0$) as pointed out in the context of excited-state nuclear dynamics\cite{karak2024reverse} and the pseudo-Jahn–Teller effect \cite{majumdar2024influence}. Despite the larger molecular size of the heptazine derivatives compared to the triangular systems, L-ADC(2) and L-CC2 provide significant speedups, enabling the calculation of their excited states. 

\begin{figure*}[!htpb]
    \centering
    \includegraphics[width=0.8\linewidth]{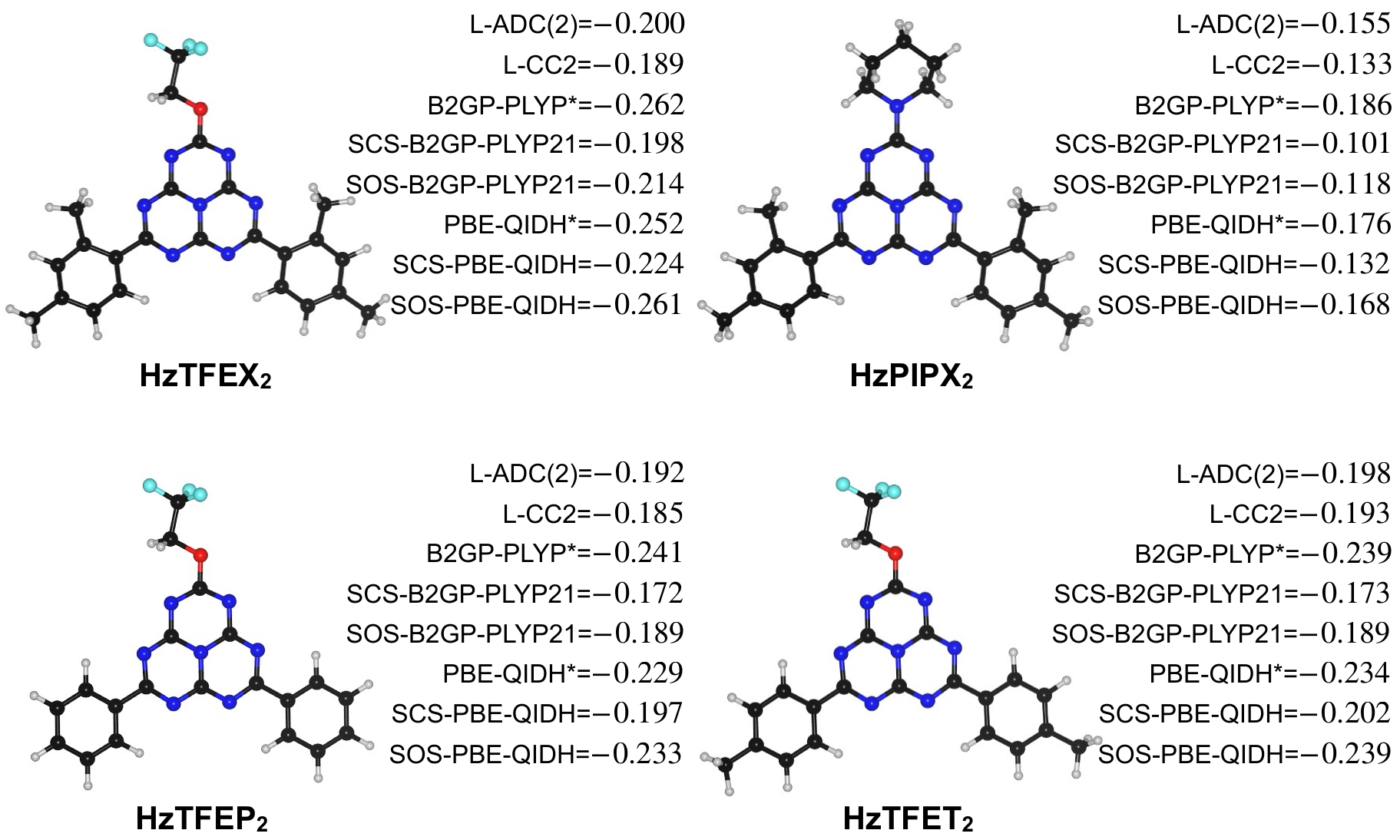}
    \caption{
Heptazine derivatives proposed as candidates for
light emitters exhibiting
delayed fluorescence from inverted singlet and 
triplet excited states (DFIST) in 
\RRef{aizawa2022delayed} are shown 
along with their computed STG values (in eV) from various methods. 
B2GP-PLYP$^*$ and PBE-QIDH$^*$ indicate that the results 
of the base methods B2GP-PLYP and PBE-QIDH
are corrected with respect to their variants with
$a_x=0.75$ and $a_c=0.55$; more details are available in the SI.
Atoms in white$|$black$|$blue$|$red$|$cyan are H$|$C$|$N$|$O$|$F. 
}
\label{fig:heptazinederivatives}
\end{figure*}

For all four systems, we performed geometry optimization at the $\omega$B97X-D3/def2TZVP level (see SI for Cartesian coordinates), followed by single-point excited-state energy calculations. 
The STGs of these four systems, obtained from various methods, are presented in Figure~\ref{fig:heptazinederivatives}. In the absence of a high-fidelity reference such as TBE for these systems, we consider L-CC2 results to be the most reliable. We used B2GP-PLYP (75,55) energies as an internal reference to scale the predictions from B2GP-PLYP (65,36) and similarly corrected PBE-QIDH (69,33) predictions using the (75,55) configuration. Notably, the application of the internal scaling does not require any additional high-level reference to identify the low-bias or low-variance domain in the parameter space of dh-DFAs. As these error metrics have been determined for the twelve benchmark systems. 
This enables a straightforward application of our internally referenced scaling procedure, which corrects systematic errors without requiring external high-level reference data.

Among the results from various dh-DFAs, SCS-PBE-QIDH shows the best agreement with L-CC2, achieving an MAD of 0.014 eV. In contrast, the bias-corrected results for B2GP-PLYP and PBE-QIDH are systematically more negative than the L-CC2 values, with MADs of 0.057 eV and 0.048 eV, respectively. While one might argue that simply shifting the results of these two DFAs could render the STGs negative, our correction is performed without requiring reference values from the reference L-CC2 values. As in the case of triangular benchmark systems (shown in FIG~\ref{fig:dataset1}), B2GP-PLYP exhibits a smaller bias compared to PBE-QIDH. Furthermore, for triangular systems, we observed that L-CC2's error relative to the TBE is consistently shifted by $-$0.013 eV, whereas the errors of the (75,55) variant of the two dh-DFAs are centered around zero (albeit with a large spread), as shown in Figure~\ref{ref:scatter_errordensity}. Consequently, when compared to L-CC2, the bias-corrected dh-DFAs are expected to exhibit a non-vanishing MSE; indeed, the MSE values for B2GP-PLYP (75,55) and PBE-QIDH (75,55) are $-$0.057 eV and $-$0.048 eV, respectively. Compared to a more accurate reference such as CC3 (which is practically challenging for the four heptazine derivatives), our bias-corrected results may show better agreement than the other methods discussed. Moreover, it is possible to identify combinations of $a_x$ and $a_c$ that yield low bias and low variance relative to L-CC2, which can be further explored for extending the bias correction scheme to larger datasets lacking high-level reference data such as TBE values.

\section{Conclusions\label{sec:conc}}
First principles excited state modeling plays a crucial role in the identification of new molecular light emitters with a negative S$_1$-T$_1$ energy gap. 
Historically, theoretical interest in such molecules existed for a long time in the context of Hund's rule violation\cite{toyota1988violation,borden1994violations,gallup1987breakdown,koseki1985violation,toyota1986violation}. 
Yet, revived interest in them has increased since their identification through first-principles modeling, not so long ago~\cite{de2019inverted,ehrmaier2019singlet}. 
Although individual S$_1$ and T$_1$ energies are challenging to predict with high precision, the error cancellation inherent in the definition of STG allows for improved accuracy, even when the magnitudes of the reported gaps are less than 0.1 eV. 
Benchmarking the STGs of twelve triangular molecules, we find that local wavefunction methods, specifically L-CC2 and L-ADC(2), deliver mean errors of approximately 15 meV with correspondingly low standard deviations, providing both exceptional accuracy and the computational speed required for high-throughput screening of medium-sized molecules such as azaphenalenes with bulky substitutions. 

In parallel, we demonstrate that double-hybrid DFT approximations, although initially characterized by high systematic errors, can be significantly improved through linear regression-based bias correction, without relying on a high-level reference. 
For example, the PBE-QIDH method in its default configuration (69\% exact exchange and 33\% MP2-level correlation) exhibits an impressively low variance (9 meV) but high mean errors. 
By adjusting the parameters to 75\% exchange and 55\% correlation, the mean error reduces to 4 meV, albeit with increased variance. 
We propose a bias correction strategy to leverage the bias-variance tradeoff inherent in the dh-DFA formalisms. 
Using a suitable low-bias method as an internal reference, the approach corrects systematic errors in low-variance models, thereby enabling accurate negative STG estimations.
In the case of four heptazine derivatives, the SCS and SOS variants of both B2GP-PLYP and PBE-QIDH yield STG predictions in better agreement with L-CC2, suggesting that these variants naturally mitigate some of the bias in standard dh-DFT. However, bias correction remains valuable for further refining low-variance models, particularly when an appropriate low-bias reference is available. 

The purpose of this study is not to advocate for developing property-specific models, but to leverage the intrinsic bias-variance tradeoff in the dh-DFA formalism for accurate STG predictions.  Open questions remain regarding the performance of these parameterized models for positive STGs and systems with charge-transfer excitations~\cite{casanova2021global}. Robust application of our strategy requires reliable reference datasets and could benefit from further bias-variance analysis using alternative information-based metrics~\cite{peverati2021fitting}. Overall, the findings of this study establish a mathematically sound framework for predictive modeling in first-principles molecular design, with promising implications for the development of novel light emitters and other applications.

\section{Supplementary Information}
i) Dataset of S$_1$, T$_1$ and S$_1$-T$_1$ energies of twelve triangular molecules
calculated with various methods and basis sets.
ii) Figures  S1--S3 present error metrics for cc-pVDZ, cc-pVTZ, and aug-cc-pVTZ basis sets;
iii) Tables S1 presents optimal $a_x$ and $a_c$ along with the associated error metrics for different energies;  
iv) Tables S2--S3 present excited state energies of heptazine derivatives; 
v) Equilibrium coordinates of heptazine derivatives.
Excited state energies of the twelve benchmark systems determined with various methods and basis sets 
are provided as CSV files in the repository: \href{https://github.com/moldis-group/triangulenes12}{https://github.com/moldis-group/triangulenes12}.

\section{Data Availability}
The data that support the findings of this study are within the article and its supplementary material.

\section{Acknowledgments}
We thank Prof. Denis Jacquemin for commenting on the ADC(2) and CC2 results of
the benchmark triangular molecules, and Dr. Marcos Casanova-P{\'a}ez for useful suggestions.  
We acknowledge the support of the 
Department of Atomic Energy, Government
of India, under Project Identification No.~RTI~4007. 
All calculations have been performed using 
the Helios computer cluster, 
which is an integral part of the MolDis 
Big Data facility, 
TIFR Hyderabad \href{http://moldis.tifrh.res.in}{(http://moldis.tifrh.res.in)}.

\section{Author Declarations}

\subsection{Author contributions}
\noindent 
{\bf AM}: 
Conceptualization (equal); 
Analysis (equal); 
Data collection (equal); 
Writing (equal).
{\bf RR}: Conceptualization (equal); 
Analysis (equal); 
Data collection (equal); 
Funding acquisition; 
Project administration and supervision; 
Resources; 
Writing (equal).

\subsection{Conflicts of Interest}
The authors have no conflicts of interest to disclose.

\section{References}
\bibliography{ref} 
\end{document}